\documentclass{article}
\usepackage{hyperref}
\usepackage{graphicx}
\usepackage[utf8]{inputenc}
\usepackage[T1]{fontenc}
\usepackage{amsmath}
\usepackage{caption}
\usepackage{authblk}
\usepackage{multirow}
\usepackage{amssymb}
\usepackage{geometry}
\usepackage{subfigure}
\usepackage{subcaption}
\usepackage{float}
\usepackage{booktabs} 
\usepackage{colortbl} 
\usepackage{xcolor} 

\title{Anisotropic anomalous diffusion in microgravity dusty plasma. Part One: Nonextensive Statistical Analysis }

\author[1]{Bradley R. Andrew\thanks{Corresponding author: bra0016@auburn.edu}}

\author[1]{Luca Guazzotto}

\author[2]{Lorin Matthews}

\author[2]{Truell Hyde}

\author[1]{E. G. Kostadinova}

\affil[1]{Physics Department, Auburn University, Auburn, Alabama, USA}
\affil[2]{Center for Astrophysics Space Physics and Engineering Research (CASPER), Baylor University, Waco, Texas, USA}

\date{}

\begin{document}
\maketitle

\begin{abstract}
Anisotropic anomalous dust diffusion in microgravity dusty plasma is investigated using experimental data from the Plasmakristall-4 (PK-4) facility on board the International Space Station. The PK-4 experiment uses video cameras to track individual dust particles, which allows the collection of large amounts of statistical information on the dust particle positions and velocities. These statistics are used to quantify anomalous dust diffusion caused by anisotropies in the plasma-mediated dust-dust interactions in PK-4. Anisotropies are caused by an externally applied polarity-switched electric field, which modifies the ion wakefields surrounding the dust grains. Video data for nine sets of pressure-current conditions are used to recover Mean Squared Displacement (\texttt{MSD}) plots after subtracting particle drift. Position and velocity histograms are fitted to Tsallis nonextensive probability distribution functions (PDFs). Both MSDs and PDFs indicate a crossover from suprathermal to L\'{e}vy diffusion in the axial direction at higher pressure conditions. In addition, increasing the pressure enhances dust thermal equilibrium, while increasing the current drives the system away from equilibrium.
\end{abstract}

\newpage

\section{Introduction }
\label{sec:introduction}

Complex (or dusty) plasmas are a collection of electrons, ions, neutral particles, and dust grains (typically micro- to nano-meter in size). Dusty plasma is ubiquitous in astrophysical and space environments, as well as in laboratory settings, both on Earth and in microgravty. Dusty plasma is a unique analogue system for the study of complex phenomena such as phase transitions, anomalous diffusion, and metastability. The particles in these systems are visible at the kinetic level, thus allowing for a reconstruction of the entire phase space. In addition, particle tracking or velocimetry techniques can be used to obtain statistically significant amount of data. Finally, dusty plasma experiments are reasonably simple to build (table-top) and highly controlled, which makes them ideal for deployment in space. Due to the complex interactions among the different charged species, dusty plasmas are observed to exhibit various waves, instabilities, and nonlinear structures \cite{shukla_survey_2001, Merlino2012}. 

\vspace{3mm}

Dusty plasmas are also ideal for studying solid-liquid phase transitions \cite{khrapak_fluid-solid_2012,APS2015,BKostadinova2023,Hariprasad2022}, electroheology \cite{ivlev_electrorheological_2010,Ivlev2011,Pustylnik2020}, strong interparticle coupling and long-range interactions \cite{tsytovich_long-range_1997,smith_dusty_2004,Correia2023}, kinetic theories and diffusion properties \cite{arshad_kinetic_2017,petrov_experimental_2005,liu_particle_2018,Feng2010}, and critical phenomena such as melting and crystallization \cite{Hariprasad2022,Feng2010,Joshi2023}, and turbulence \cite{kostadinova_fractional_2021,Sharma2024,Choudhary2024}. As they exhibit many-body effects, dusty plasmas are useful analogue systems for the study of complex systems such as condensed matter \cite{murillo_strongly_2004,kostadinova_fractional_2021} and smart materials \cite{ivlev_first_2008,ivlev_electrorheological_2010,dietz_phase_2021}. Many other physical aspects and applications of dusty plasma have been summarized in several recent overview papers\cite{Beckers2023,choudhary_perspective_2021,Merlino2021}. With all the characteristics described above, dusty plasmas are ideal for testing new analytical models, especially nonequilibrium statistical mechanics anomalous diffusion, and stochasticity \cite{kostadinova_fractional_2021,kostadinova_delocalization_2017,kostadinova_physical_2016,kostadinova_transport_2018,liu_non-gaussian_2008,liu_particle_2018}. In part two of this paper, we will be analyzing a very recent spectral model using an Anderson Type Hamiltonian with a long-range Fractional Laplacian operator \cite{kostadinova_fractional_2021,kostadinova_delocalization_2017,kostadinova_physical_2016,kostadinova_transport_2018} and relating it to the same data set presented here.

\vspace{3mm}

Here we investigate dusty plasma experiments conducted in the Plasmakristall-4 (PK-4) facility on board the International Space Station, where the microgravity environment allows to neglect gravity and confinement forces, thus, focusing on plasma-mediated dust-dust interactions. Recent studies using the PK-4 facility \cite{pustylnik_plasmakristall-4_2016} have investigated various dynamical phenomena, including dust ionization waves \cite{Naumkin2021,Zhukhovitskii2022,Pustylnik2022}, ion density waves \cite{Mendoza2024}, dust acoustic waves \cite{Goree2020},unsteady shear flows and flow patterns fluctuations \cite{Liu2021}, and lunar atmosphere dust \cite{freeman2024lunar}. Additionally, analysis of PK-4 data has inspired a breadth of numerical studies, including particle-in-cell simulation of PK-4 predicting the formation of ionization waves \cite{Hartmann2020} and molecular dynamics simulations of dust and ions investigating how such ionization waves can cause anisotropies in the ion wakefields around the dust grains \cite{Vermillion2022,Vermillion2023}. Finally, non-Mazwellian distributions of the dust velocities have been observed in PK-4 experiments with RF discharge configuration \cite{liu_particle_2018}. Here we discuss the observation of non-Maxwellian distribution functions in PK-4 experiments with a pure DC discharge, where a polarity-switching external electric field causes anisotropies on the dust-dust interaction potential and the resulting dust diffusion. 

\vspace{3mm}

The PK-4 experiment uses video cameras to track individual dust particles, which allows for obtaining large amounts of statistical information on the dust particles positions and velocities. Previous studies of dusty plasma with the PK-4 experiment have shown velocity distribution functions (VDFs) that were non-Maxwellian, had high-energy tails, and displayed anomalous diffusion. Anomalous diffusion is a microscopic process which leads to a mean squared displacement (MSD) that grows non-linearly with time $MSD\propto \tau^\alpha$, where $\tau$ is time delay. If the MSD growth is faster than linear with time, $\alpha>1$, the particles are superdiffusive, while growth rate slower than linear, $\alpha<1$, indicates subdiffusion. Anomalous diffusion and corresponding non-linear MSDs have been observed in biological molecular transport \cite{tarantino_tnf_2014}, L\'{e}vy flight movement patterns of living organisms \cite{reynolds_liberating_2015}, condensed matter physics \cite{benhamou_lecture_2018}, and dusty plasmas \cite{feng_identifying_2010,liu_non-gaussian_2008}. A review of classical and anomalous diffusion across many fields can be found in \cite{oliveira_anomalous_2019}. One of the goals of the present paper is to promote the use of Tsallis nonextensive statistics in dusty plasma analysis, since it provides greater clarity to the classification and study of nonequilibrium systems, non-Maxwellian distributions, and the resulting anomalous diffusion.

\vspace{3mm}

Here we analyze anomalous diffusion and thermodynamic nonequilibrium properties by constructing MSD plots and histograms of dust displacements and velocities. Nonextensive (Tsallis) statistics \cite{daniels_defect_2004,reynolds_rotational_2004,Urgur_aging_2005,abul-magd_nonextensive_2005,ivanova_dynamical_2007,douglas_tunable_2006,bediaga_nonextensive_2000} is used to quantify how dust PDFs deviate from a Gaussian/Maxwellian distribution and to identify the sub-regime of anomalous diffusion that best describe each dataset. Nonextensive statistics is a formulation of statistical mechanics where entropies are nonadditive (nonextensive). Nonextensive statistics has been used in plasma physics to better understand solar wind turbulence and dynamics \cite{karakatsanis_tsallis_2013,pavlos_tsallis_2012}, plasma waves \cite{Summers1991,Liu2009,Bilal2023}, collisions \cite{wang_collision_2021}, dusty plasmas \cite{Gong2012,LiuDu2009}, including analysis of previous PK-4 experiments using combined DC-RF discharge \cite{liu_non-gaussian_2008,liu_particle_2018}. We build on these previous studies by analyzing PK-4 data from pure DC discharge experiments where the dust clouds were shown to exhibit strongly non-isotropic filamentary structure.

\vspace{3mm}

 We investigate nine pressure-current datasets from PK-4 experiments conducted in pure DC neon discharge. The negatively-charged dust particles are kept stationary in the field of view of the particle observation cameras by switching the polarity of an externally-applied electric field. As the frequency of the polarity switching (500Hz) is higher than the typical dust response frequency, the dust experiences net zero force due to the electric field. This, however, results in an anisotropy in the ion wakefields surrounding the dust, which in turn causes anisotropic dust diffusion. Particle tracking techniques \cite{schindelin_fiji_2012} were used to obtain the dust positions and velocities from video data. The dust MSD, the displacement histograms, and the velocity histograms were reconstructed using the open-sourced \texttt{@msdanalyzer} code \cite{tarantino_tnf_2014}. Table \ref{tab:exp_params} below provides a summary of the plasma conditions and dust density for each analyzed case.

\begin{table}[H]
    \centering
    \begin{tabular}{|c|c|c|c|c|c|c|c|c|c|} 
    \hline  
         Data Set & 1 & 2 & 3 & 4 & 5 & 6 & 7 & 8 & 9 \\ 
    \hline  
         $P$ [Pa] & 28.5 & 28.5 & 28.5 & 46.1 & 46.1 & 46.1 & 70.5 & 70.5 & 70.5 \\ 
    \hline
         $I$ [mA] & 0.35 & 0.7 & 1 & 0.35 & 0.7 & 1 & 0.35 & 0.7 & 1 \\ 
    \hline
         $n$ [mm$^{-3}$] & 81.8 & 88 & 85.3 & 123.6 & 93.4 & 93.3 & 55.1 & 93.3 & 69.3 \\ 
    \hline 
    \end{tabular}
    \caption{Pressure, current, and dust density for each examined data set from PK-4 experiments.}
    \label{tab:exp_params}
\end{table}

Fits to the MSDs plots reveal a non-linear relation with time delay $\tau^\alpha$, which is indicative of anomalous diffusion. Fits to both the position and velocity histograms for the direction along the external electric field are best described by a q-Gaussian distribution function, where the non-extensive exponent $q$ quantifies the 'tailedness' of a distribution deviates from a standard Gaussian or a Maxwellian one. To determine the diffusion sub-regime, we use scaling relations from literature to compare the nonextensive exponent $q_p$ of position distributions (found from fits to displacement histograms) against the MSD exponent $\alpha$. The dust displacements and velocities in the direction perpendicular to the external electric field are best described by a Bi-q-Gaussian distribution, which is a sum of two q-Gaussian distributions. This suggests that the system has two distinct thermodynamic populations. As the neutral gas pressure in dusty plasma mediates the dust-neutral collisions, increasing pressure in these experiments acts as decreasing temperature, which is why the observed thermodynamics is sensitive to pressure changes. This was also seen by \cite{Baylor2024}, which conducted a structural analysis of the same set of experiments using pair correlation function techniques. The remainder of this paper is dedicated to the qualitative and quantitative description of the observed anomalous diffusion, anisotropies, and equilibrium properties of dusty plasma in PK-4.

\vspace{3mm}

The remainder of this paper is organized as follows. An overview of the experimental setup used for the PK-4 experiment is provided in (\autoref{sec:Experiment}). A summary of nonextensive statistics and its application to anomalous diffusion is given in (\autoref{sec:Theory}).  The statistical and fitting tools used for the analysis are described in (\autoref{sec:msdanalyzer}). A summary of results is presented in (\autoref{sec:Analysis}) followed by a discussion in (\autoref{sec:Discussion}). Conclusions and future work are outlined in (\autoref{sec:Conclusions}).

\section{Experimental Setup}
\label{sec:Experiment}
\begin{figure}[ht]
    \centering
    \includegraphics[width=80mm]{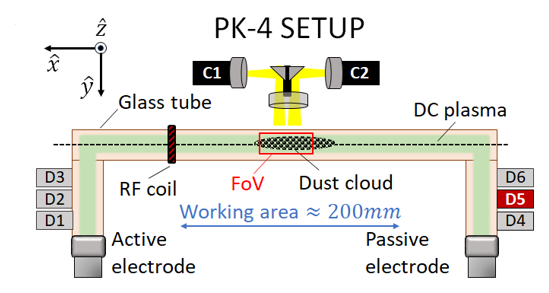}
    \caption{PK-4 Experimental Setup. Dust moves into camera FoV.}    \label{fig:PK4}
\end{figure}

Here we briefly discuss the PK-4 experimental apparatus \cite{pustylnik_plasmakristall-4_2016}, and the specifics of the Campaign 7 (C7) data that we use for the analysis. The core of PK-4 comprises an integrated baseplate housing the diverse components, including a glass plasma chamber with electrodes and microparticle dispensers (injecting Melamine-Formaldehyde spheres), vacuum and gas supply systems, plasma generation and diagnostic tools, microparticle manipulation devices, a microparticle observation system with cameras, and an illumination laser, see Figure \ref{fig:PK4}. The main vacuum vessel is a cylindrical glass chamber, where plasma can be created using a dc discharge power supply. Polarity switching of the dc current at different frequencies and variable duty cycles can be used to transport the dust particles (using an asymmetric duty cycle) and capture them in the cameras field of view (using a symmetric duty cycle). Several microparticle manipulation options and on-board plasma diagnostics, such as a plasma glow observation system and a mini spectrometer, are also available. Here we label the axial direction in  PK-4 (x direction in Figure \ref{fig:PK4}) as $\|$ since it is parallel to the direction of the dc electric field. The radial, or cross-field, direction in the camera's field of view (z direction in Figure \ref{fig:PK4}) will be labeled $\perp$.

\vspace{3mm}

The DC discharge plasma is generated by two electrodes within a $\pi$-shaped glass chamber. A custom-made bipolar high-voltage (HV) power supply serves as a current source, providing a stabilized output current up to 3.1 mA at a maximal overall voltage of 2.7 kV. The current is regulated on the active side, with return current measurement on the passive side. Steady-state deviations from the set current value remain below 5$\%$. When symmetric duty cycle is used with a fast polarity switching of the current (here 500 Hz), the dust microparticles are unresponsive as dust response frequency is close to 10 Hz. This results in overall stationary negatively charged microparticles suspended within a slowly 'sloshing' stream of ions that forms anisotropic ion wakefields surrounding the dust grains. The Particle Observation (PO) system facilitates microparticle imaging, employing a 532 nm diode laser and two PO cameras with CCD chips of 1600 × 1200 pixels. The cameras are movable and can cover the entire volume of the working area. 

\vspace{3mm}

The PK-4 Campaign 7 experiments discussed here were conducted on July 26, 2019. Our analysis uses the video data from nine sets of pressure-current conditions (see Table \ref{tab:exp_params}). The current polarity switching was 500 Hz with a duty cycle of $50\%$, and microparticle size of $3.38 \mu m$ diameter. In each case, the dust cloud was allowed to settle for ~50 s, after which a scan of the laser sheet was performed through the dust cloud (along the y-axis in Figure \ref{fig:PK4}). These y-scans allow for obtaining information on the 3D structure of the cloud. The statistical analysis presented here uses particle tracking data obtained for a period in which the dust cloud has settled and that the cameras and laser sheet are focused on the mid-plane of the cloud. We use the dust particle positions and velocities to study the probability distribution functions, anomalous diffusion, temperature, and nonequilibrium properties.

\section{Statistical Approach to Anomalous Diffusion}
\label{sec:Theory}
\subsection{Diffusion}

In the absence of long-range interactions or correlations, the particle diffusion can be described by Brownian motion and the model differential equation is the well-known diffusion equation

\begin{equation}
    \frac{\partial p(x,t)}{\partial t}= \Delta ( D p(x,t)),
    \label{eq:normaldiff}
\end{equation}

where $\Delta=\frac{\partial^2}{\partial x^2}$ is the Laplacian operator, D is the diffusion constant, and p(x,t) is the distribution function. Equation (\ref{eq:normaldiff}) is also the linear Fokker-Planck equation with no drift. The one-dimensional solution to the diffusion equation has a Gaussion distribution functional form given by

\begin{equation}
    p(x,t)=\frac{1}{\sqrt{4\pi Dt}} e^{-\frac{(x-x_0)^2}{4Dt}}.
    \label{eq:diffusioneq}
\end{equation}

The mean squared displacement (MSD) is the second moment of this equation: $\langle(x-x_0)^2\rangle\equiv \int (x-x_0)^2 p(x,t)dx=Dt$. As can be seen, the classical diffusion equation yields MSD that grows linearly with time. Assuming discrete particle positions, the MSD expression for N particles can be generalized in two or three dimensions to

\begin{equation}
    \langle |r(t)-r_0|^2\rangle = \frac{1}{N}\sum_{i=1}^N |r^{(i)}(t)-r^{(i)}(t=0)|^2 = 2dDt^\alpha   \hspace{4mm}  \alpha \geq 0, 
    \label{eq:genmsd}
\end{equation}

where $d$ is the dimension and $\alpha$ is an exponent that quantifies deviations from a linear dependence in time (i.e., the presence of anomalous diffusion. When $\alpha=1$ (linear MSD), the diffusion is classical, while $\alpha\neq 1$ implies anomalous diffusion. While anomalous diffusion is ubiquitous in nature  \cite{tarantino_tnf_2014, reynolds_liberating_2015, benhamou_lecture_2018,feng_identifying_2010,liu_non-gaussian_2008,oliveira_anomalous_2019}, its mathematical description is challenging. One way to study anomalous diffusion is with a nonlinear diffusion equation, the nonlinear Fokker-Planck differential equation, which has solutions including Porous Medium equations (see page 249 of \cite{frank_nonlinear_2005}) and Tsallis q-Gaussian distributions \cite{tsallis_introduction_2009} (also known as Kappa distribution or Student's t distribution). Anomalous diffusion may also be studied by use of fractional spatial derivatives \cite{kostadinova_delocalization_2017,kostadinova_transport_2018,padgett_anomalous_2020,kostadinova_fractional_2021} which may have solutions such as L\'{e}vy distributions. We note that it is not well understood if universal scaling relations exist 
 that link the different proposed analytical formulations for anomalous diffusion. While there are some direct relations between the nonlinear Fokker-Planck equations and the q-Gaussian solutions, their relation to fractional differential equations is not straightforward even though they can both describe similar regimes of anomalous diffusion.

\subsection{The Nonlinear Diffusion Equation and Tsallis' Statistics}

The nonlinear Fokker-Planck equation with no drift

\begin{equation}
\frac{\partial p(x,t)}{\partial t}=D\frac{\partial^2 [p(x,t)]^\nu}{\partial x^2}
\label{eq:nonlindiff}
\end{equation}

has the following explicit solution derived in \cite{frank_nonlinear_2005}, page 249

\begin{equation}
p(x,t) = A_{\nu}^{-1} \left[ 2D(\nu+\nu^2)(t-t_0) \right]^{\frac{-1}{1+\nu}} 
\cdot \left[ 1 + \frac{(1-\nu) (x-x_0 )^2}{\left( (2D(\nu+\nu^2)(t-t_0)A_{\nu}^{1-\nu})^{\frac{2}{1+\nu}} \right)} \right]^{\frac{-1}{1-\nu}}.
\label{eq:Dq}
\end{equation}

The expression can be rewritten using Tsallis nonextensive $q$ using the substitution $\nu \equiv 2-q$, which yields

\begin{equation}
p(x,t) = A_{q}^{-1} \left[ 2D(6-5q+q^2)(t-t_0) \right]^{\frac{-1}{3-q}} \left[ 1 + \frac{(q-1) (x-x_0 )^2}{\left( (2D(6-5q+q^2)(t-t_0)A_{q}^{q-1})^{\frac{2}{3-q}} \right)}\right]^{\frac{-1}{q-1}}
\end{equation}

 This is similar to the expression provided in \cite{tsallis_introduction_2009}, where $A_q$ is a normalization factor given by

\begin{equation}
A_{q} = 
\begin{cases}
\sqrt{\frac{\pi}{(1-q)}}\frac{\Gamma(\frac{2-q}{1-q})}{\Gamma(\frac{q}{2-2q})} & \text{if } q<1 \\                                  
\sqrt{\pi}  & \text{if } q=1 \\
\sqrt{\frac{\pi}{(q-1)}} \frac{\Gamma(\frac{3-q}{2q-2})}{\Gamma(\frac{q}{q-1})} & \text{if } 1<q<3.
\end{cases}
\label{eq:qnorm}
\end{equation}

When $q=1$ and $t_0 = 0$ this recovers equation \ref{eq:diffusioneq}. The parameter $q$ can also be related to the distribution kurtosis $k$, through the following relationship: $k = \frac{15-9q}{7-5q}$, for $q<7/5$ \cite{q2kurtSalah} where $k=Kurtosis=\frac{\langle x^4\rangle}{\langle x^2\rangle^2}$. This can be used to quantify the deviations from a Gaussian distribution and the related anomalous diffusion (e.g., as discussed in \cite{Ghannad2021}). The primary interest of the present study is the range $1<q<3$ where the distributions exhibit leptokurtic or ‘fat-tail’ behavior. This regime corresponds to superdiffusive transport ($\alpha >1$), though we will also discuss subdiffusive cases where $q<1$ or $\alpha<1$ (see Figure 2 for example distributions representing each regime). Anomalous diffusion is described by taking the second moment of Eq. \ref{eq:Dq}, which yields a non-linear relation between the MSD and time

\begin{equation}
\langle (x - x_0)^2 \rangle = B_q (D \tau)^{\frac{2}{3-q}}, \quad \tau = t - t_0
\label{eq:alpha2q}
\end{equation}

where

\begin{equation}
B_q = \frac{2^{\frac{1-q}{q-3}} \pi^{\frac{3(1-q)}{2(q-3)}} 
(q-1)^{\frac{q}{2(q-3)}}
\Gamma\left(\frac{4-3q}{2q-2}\right)}
{(q-3)^{\frac{2}{q-3}} (q-2)^{\frac{2}{q-3}} 
\Gamma\left(\frac{1}{q-1}\right) 
\left( \frac{\Gamma\left(\frac{2-q}{2q-2}\right)}
{\Gamma\left(\frac{q}{q-1}\right)} \right)^{\frac{4q - 6}{q-3}}}
\end{equation}

Nonextensive statistics was proposed by Tsallis \cite{tsallis_possible_1988} and is fully described in his textbook \cite{tsallis_introduction_2009}. The probability distribution function in this formulation is called a q-Gaussian, where the parameter $q$ quantifies the nonextensivity of the system. The nonextensivity feature of the formulation makes it independent of initial conditions and suitable for modeling many-body complex systems with long-range interactions, such as plasma. Systems with this nonextensive behavior no longer have additive entropies, $S_q(A+B)\neq S_q(A) + S_q(B)$, and the different microstates are not in equilibrium. Therefore, quantifying the nonextensivity is needed to make rigorous statistical claims and understand equilibrium properties of systems with long-range interactions and correlations. Thus, $q$ can be understood as a parameter quantifying the strength of correlations or non-local interactions causing the system to move away from equilibrium. The nonextensive parameter $q$ can be found by extremizing the Tsallis entropy $S_q$ with Lagrange constraints, such as normalization and moment of the distribution. Essentially, it can be thought of as the nonlinear form of the Maxwellian distribution, with normalization $A_q$, kinetic energies $E_i$, potential $U$, and inverse temperature $\beta_q$. This results in the following probability distribution function, known as the q-Gaussian

\begin{equation}
p_i=\frac{\beta_q}{A_q} e_q^{(-\beta_q(E_i-U_q))}= \frac{\beta_q}{A_q} [1-(1-q)\beta_q(E_i-U_q)]^{\frac{1}{(1-q)}}
\label{eq:qdist}
\end{equation}

Note that in the limit where $q\rightarrow1$, the Maxwellian distribution is recovered. This is accomplished by letting the potential energy $U_q=0$, $\beta_q=1/(v_{qth}^2)$, where $v_{qth}$ is the q-Gaussian thermal velocity, and $E=\frac{1}{2}mv^2$, recalling the fact that $e^x = \lim_{n \to \infty} \left(1 + \frac{x}{n}\right)^n$ and letting $n=\frac{1}{1-q}$

\begin{equation}
\lim_{q\rightarrow 1} \frac{1}{v_{qth} A_q} [1-(1-q)\frac{mv^2}{2 v_{qth}^2}]^{\frac{1}{(1-q)}} \rightarrow \sqrt{\frac{m}{2\pi v_{th}^2}}*e^{-\frac{v^2}{2m v_{th}^2}}
.\label{eq:qdist2max}
\end{equation}

The relationship between the Maxwellian kinetic temperature $T_M$ and the q-Gaussian kinetic temperature $T_q$ was discussed in  \cite{Bilal2023} and is given by

\begin{equation}
    T_q(\frac{5q-3}{2})=T_M.
    \label{eq:qvariance}
\end{equation}

Eq. \ref{eq:qvariance} relates the q-Gaussian variance fits (i.e. $\sigma^2_q$, $D_q$, $v_{qth}^2$, or $T_q$) to the corresponding Gaussian or Maxwellian value. This step is necessary as the kurtosis of a distribution affects the variance, which should be taken into account to find a more appropriate variance value.

\begin{figure}[H]
    \centering
    \includegraphics[width=80mm]{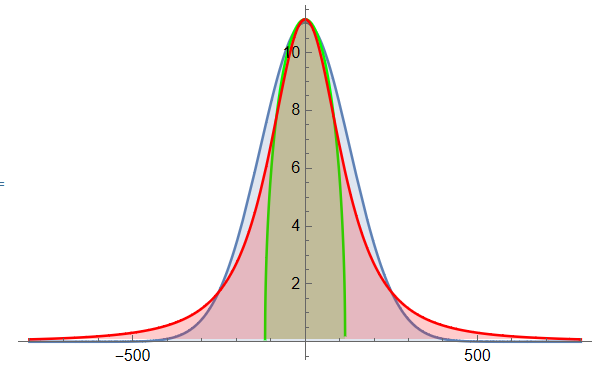}
    \caption{Analytical plot of the q-Gaussian distribution for three cases. $q=-1$(Green), $q=1$(Blue), and $q=5/3$(Red).}    
    \label{fig:qgaussian}
\end{figure}

Figure \ref{fig:qgaussian} shows different q-Gaussian distributions scaled to the same height to exemplify how the distribution shape deviates from a Gaussian ($q=1$) for different values of the nonextensive parameter $q$. Equation \ref{eq:Dq} and Eq. \ref{eq:qdist} have the same functional form, but have different variables, coefficients, and exponents as the former is in position space, while the latter is in velocity space. In this paper, we distinguish between the two by using $q_p$ and $q_v$, respectively. The variances of these distributions are finite for $q<5/3$ but diverge for $5/3\leq q<3$. This means that, as $N\rightarrow \infty$, the distribution approaches a Gaussian for $q<5/3$, while, for $5/3\leq q<3$, it approaches a L\'{e}vy distribution. For the position displacement distributions, $q_p>5/3$ the diffusion is a L\'{e}vy process which yields large 'jumps' or L\'{e}vy flights. Since we utilize the notation $q_v$, we call a process 'energetic' or 'suprathermal' if $q_v>\frac{4+d}{2+d}$, where $d$ is the dimensionality. Thus, for $d=1$,  $q>\frac{5}{3}$ there are 'large jumps' in velocity space, as expected for a L\'{e}vy process. For $q<\frac{5}{3}$ (implying the mathematical variance is finite), there are no L\'{e}vy processes or L\'{e}vy flights. In part two of this paper, "Spectral Analysis" we will go into more detail on this topic and the use of the one-dimensional Fractional Laplacian to model the different sub-regimes of anomalous diffusion. 

\section{Analysis Techniques}
\label{sec:msdanalyzer}

In this section, we discuss the techniques used for the statistical analysis of PK-4 data. Those include particle tracking, drift subtraction, and techniques for fitting functions to MSD plots and velocity/displacement histograms reconstructed from experimental data. The analysis codes used in this work are all open sourced and can be found at https://github.com/IPL-Physics/Open-Source.

\subsection{Particle Tracking and Drift Subtraction}

Here we consider PK-4 particle observation camera videos in which the laser sheet was fixed in the central region of the cloud for extended periods of time (about $20~s$). The camera frame rate used was $71~fps$ resulting in a $0.014~s$ time step. The region of interest in the videos was a rectangular section in the center of each cloud with size $14~mm$ by $2~mm$. The typical number of particles detected per frame was $~470$. This yields statistically significant datasets with $20k$ data points or more in each set. The particle positions in the xz-plane were obtained using the open source particle tracking Mosaic Suite plugin of Fiji, which is a distribution of ImageJ \cite{mosaic_fiji_2013}. The dust positions were converted from pixels to $\mu$m assuming a pixel resolution of $14.20~\mathrm{\mu m}$ \cite{pustylnik_plasmakristall-4_2016}. The particle trajectories imported from ImageJ include trajectories that do not start at the same time and do not last for the same duration. As shorter tracks can correspond to particles that move in and out of the plane illuminated by the laser sheet, we discarded tracks where the particles appeare in fewer than 10 frames (or $0.14~s$). The remaining trajectory data were used as inputs in the \texttt{@msdanalyzer} code to subtract drift, compute MSDs, and construct histograms of particle displacements and velocities. The nine pressure-current cases considered here use identical particle tracking datasets as those in \cite{Baylor2024}, which investigated the anisotropic structure of the PK-4 clouds using pair correlation analysis. Thus, we expect that the results obtained here and in \cite{Baylor2024} are directly comparable. 

\vspace{3mm}

The open-sourced code \texttt{@msdanalyzer} \cite{tarantino_tnf_2014} was used to analyze the data. The developers of \texttt{@msdanalyzer} also developed the Mosaic Suite in Fiji, providing good compatibility between the code and the particle tracking data to construct the MSD (with options to provide the standard deviation and check for localization error) and to create the histograms of dust displacements and velocities. The \texttt{@msdanalyzer} code was also used to calculate the velocity autocorrelation, detect directed motion, and subtract drift. Those steps were necessary to ensure that the statistical analysis is performed on the diffusion portion of the dust motion, subtracting convective motion. In addition, motion along the direction of the external electric field $E$ (the x-axis or the $\parallel-$ direction) is considered separately from motion across the the direction of $E$ (the z-axis or the $\perp-$ direction). This was necessary due to the anisotropic coupling of the dust particles discussed in \cite{Baylor2024}.

\vspace{3mm}

\begin{figure}[ht] 
    \centering
    \includegraphics[width=160mm,height=80mm]{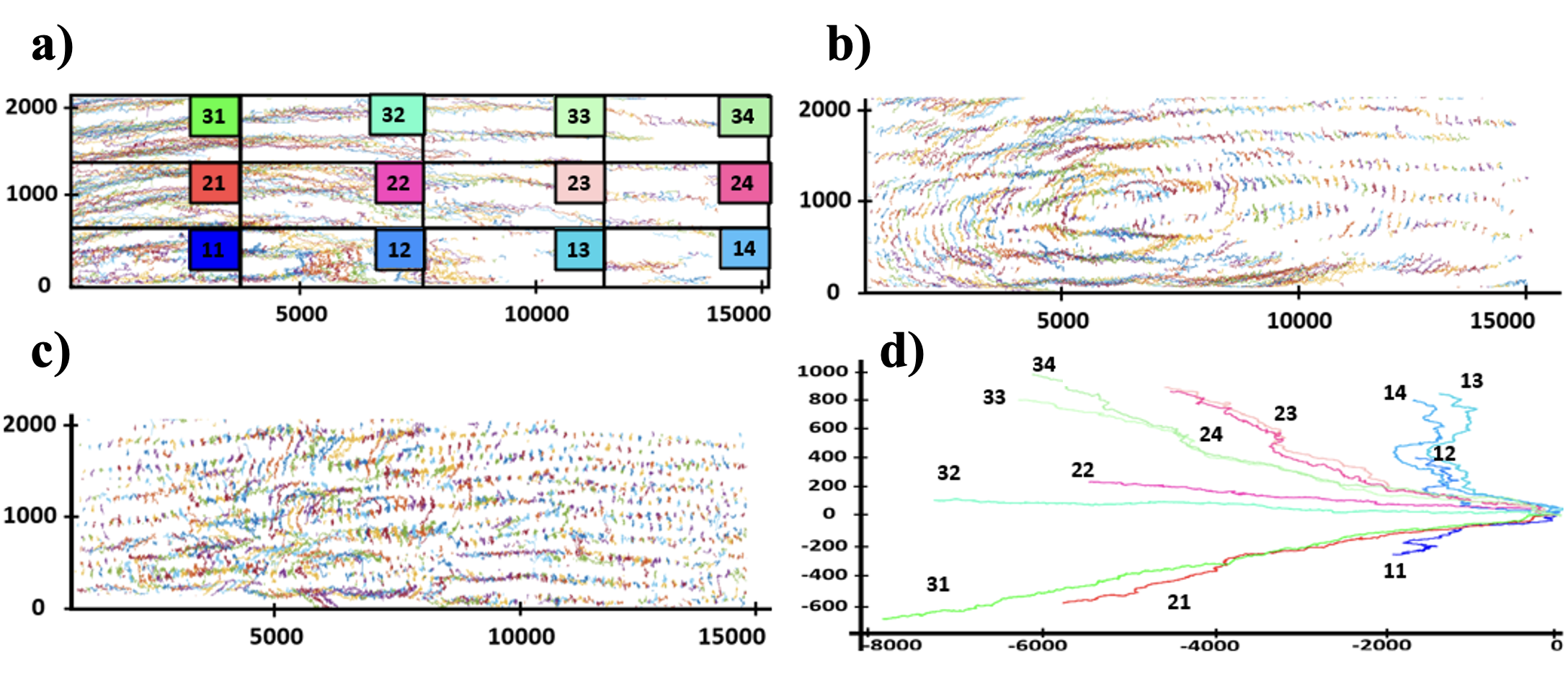}
    \caption{a) Dust particle tracks ($\mu m$) found using \texttt{@msdanalyzer} broken into 12 domains. b) Dust tracks ($\mu m$) after normal drift subtraction. c) Dust tracks ($\mu m$) after nonhomogeneous drift subtraction (NHDS). d) Average drift trajectories found in each domain in A). Experimental conditions were 70 Pa 0.35 mA.} 
    \label{fig:nhds}
\end{figure}

Each PK-4 dataset used in our analysis was extracted from a large field of view, encompassing a large portion of the dust cloud. As a result, we could break up the field of view in to smaller sections and calculate different drifts in different domains or sections of the cloud. This was necessary as nonhomogeneous drift was observed for several datasets. To address this issue, a method of nonhomogeneous drift subtraction (NHDS) was developed. In this method, the large field of view is separated into 12 smaller regions, as shown in Figure \ref{fig:nhds} A, and homogeneous drift subtraction is performed for each region. The number of smaller regions was selected to balance accuracy in drift subtraction and reasonable computation times. The NHDS method was validated against cases where the normal homogeneous drift subtraction worked, such as the case at 70 Pa pressure and 0.7 mA dc current. We found the two methods produced the same results, with similar velocity auto-correlation and the total drift plots.

\vspace{3mm}

Figure \ref{fig:nhds} shows the case where the most pronounced nonhomogeneous drift was observed for 70 Pa pressure and 0.35 mA dc current. As can be seen in Figure \ref{fig:nhds}  B, in the presence of nonhomogeneous drift, applying the drift subtraction to the large region of interest produces particle trajectories that exhibit some coordinated motion (instead of showing random diffusive motion). The NHDS method seems to work better as shown in \ref{fig:nhds} C, though some small areas still exhibit coordinated motion instead of the desired diffusive behavior. The drifts for each smaller domain region are shown in \ref{fig:nhds} D. While the total drift simply implied that the dust particles were linearly drifting to the left, the domain drifts show a spread of drift directions. 

\begin{figure}[H] 
    \centering
    \includegraphics[width=120mm,height=60mm]{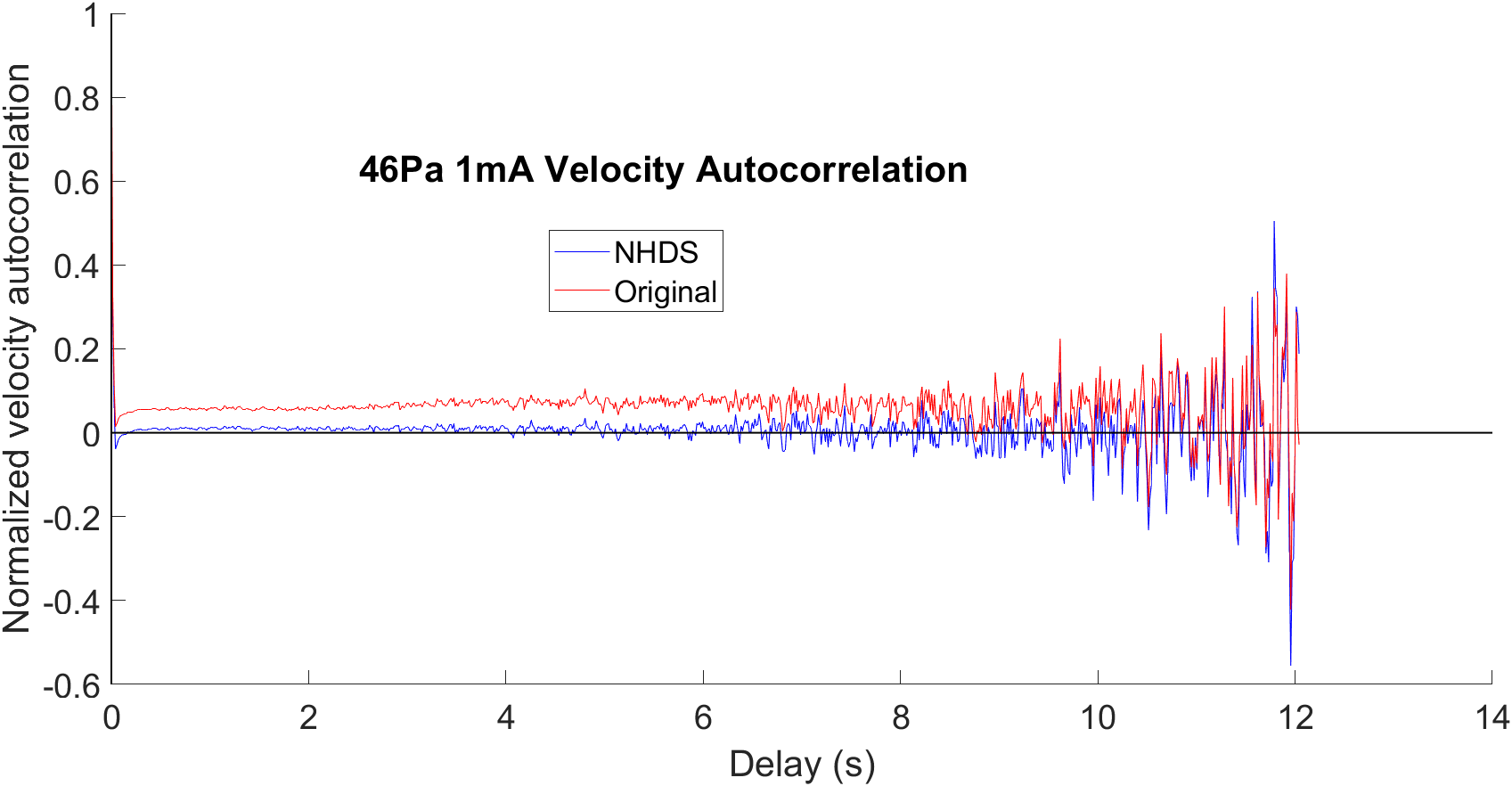}
    \caption{Velocity Autocorrelation of the 70 Pa 0.35 mA case before (red) and after NHDS (blue).}
    \label{fig:autocorr}
\end{figure}

As a quantitative check of the NHDS application of drift correction for the 70 Pa 0.35 mA case, we calculated the velocity autocorrelation of all domains before and after, shown in Figure \ref{fig:autocorr}. While the velocity autocorrelation before subtraction (red) showed a nonzero value, after the NHDS subtraction (blue), the total autocorrelation dropped to zero. Using these drift-subtracted particle trajectories, we proceeded to find the values of $\alpha$ from a fit to the MHDS plots and the values of $q_p$ and $q_v$ from fits to the position and velocity histograms. The values of $\alpha$ obtained from fits to the MSD plots at time delay of five seconds are very similar for two drift subtraction methods. However, the application of the NHDS method decreased the magnitude of th measured MSD (reflected in the vertical axis) by an order of magnitude.

\subsection{Fitting Techniques}

In-house codes were developed to obtain nonlinear fits to the MSDs at different time delay intervals $\tau$ and distribution fits to the histograms. Fits to the displacement and velocity histograms were obtained using a Maxwellian (Eq. \ref{eq:qdist2max}), a single q-Gaussian, and a Bi-q-Gaussian distributions of the form

\begin{equation}
    \frac{A}{\sqrt{\pi v_{th}^2}}  \left[1+(q_{v}-1)\frac{(v_-v_{0})^2 }{v_{th}^2} \right]^{\frac{-1}{q_{v}-1}}  \hspace{8mm}   1\leq q_{v}<3
    \label{eq:qvx}
\end{equation}

\vspace{4mm}

\begin{equation}
\begin{aligned}
    \frac{A}{\sqrt{\pi v_{th1}^2}}  \left[1+(q_{v1}-1)\frac{(v-v_{0})^2 }{v_{th1}^2} \right]^{\frac{-1}{q_{v1}-1}} \hspace{10mm} \\ 
    +\frac{A}{\sqrt{\pi v_{th2}^2}}  \left[1+(q_{v2}-1)\frac{(v-v_{0})^2 }{v_{th2}^2} \right]^{\frac{-1}{q_{v2}-1}} 
\end{aligned}
 \hspace{8mm}1\leq q_{v1,v2} <3.	
 \label{eq:qv}
\end{equation}

Here $A,v_{th},v_0$, and $q$ are the fitting parameters for the velocity histograms. The fits to the position histograms used the same distributions where $v_{th}$ was replaced by $D$ and  $(v_r-v_{r0})^2$ was replaced by $(r(\tau)-r_0)^2$. To minimize fitting errors, the normalization in Eq. \ref{eq:qnorm} was set to $1$ and the fitting function from Eq. \ref{eq:qdist} from \cite{tsallis_introduction_2009} was used. Dropping this normalization has a negligible effect. For all cases, it was found that the axial component of the velocities (displacements) is best approximated by a single q-Gaussian,  while the radial component was best approximated by a Bi-q-Gaussian. The value of $q$ found from the position and velocity histogram fits are denoted as $q_p$ and $q_v$, respectively. Data obtained from particle displacements/velocities along the axial direction is labeled by a subscript $\parallel$, while cross-field direction data have a subscript $\perp$. The kinetic temperature was calculated from the thermal velocity parameters, while it is defined as

\begin{equation}
        T = \frac{m}{3k_B} \int (v - v_0)^2 f(v) dv = \frac{m}{3k_B} \langle(v - v_0)^2\rangle.
\end{equation}

This still requires the distribution curve $f(v)$ which we are not assuming to be Maxwellian but finding from the histogram. It is therefore faster to calculate temperature from the fitted variance which is the thermal velocity than to integrate the function to find the second moment of the distribution which is the variance or Temperature. Also, the integral does not converge for $q>5/3$ as discussed earlier, however for $q<5/3$ both methods were compared and result was a small difference of a few Kelvin.

\newpage

\section{Results}
\label{sec:Analysis}

In this section, the MSD exponents $\alpha$ and nonextensive parameters $q_p$ and $q_v$ are given subscripts corresponding to the directions parallel and perpendicular to the electric field in the PK-4 experiment. The cross-field position histograms are best approximated by Bi-q-Gaussian with a mostly Gaussian "sub-population" ($q_{\perp1}\approx 1$)  and a 'halo-tail' ($q_{\perp2}>1$). Table \ref{tb:fitparams} provides the fitted parameters $\alpha$, $q_p$, and $q_v$ quantifying anomalous diffusion and nonequilibrium in the parallel and perpendicular direction for the different pressure-current cases.

\begin{table}[h]
    \centering
    \begin{tabular}{|c|c|c|c|c|c|c|c|c|c|} 
        \hline
        \multicolumn{2}{|c|}{Pressure-Current} & $\alpha_{\parallel}$ & $\alpha_{\perp}$ & $q_{p\parallel}$ & $q_{p\perp1}$ & $q_{p\perp2}$ & $q_{v\parallel}$ & $q_{v\perp1}$ & $q_{v\perp2}$ \\ 
        \hline
        \multirow{3}{*}{70 Pa} & 1 mA & 2.01 & 1.05 & 1.81 & 1.40 & 1.60 & 1.14 & 1.20 & 2.21 \\ 
        \cline{2-10}
                               & 0.7 mA & 1.68 & 1.13 & 1.70 & 0.98 & 1.25 & 1.13 & 1.30 & 1.63 \\ 
        \cline{2-10}
                               & 0.35 mA & 2.05 & 1.45 & 1.55 & 0.96 & 1.38 & 1.17 & 1.23 & 1.38 \\ 
        \specialrule{1.5pt}{0pt}{0pt} 
        \multirow{3}{*}{46 Pa} & 1 mA & 2.40 & 0.925 & 1.44 & 0.98 & 1.27 & 1.31 & 1.12 & 2.12 \\ 
        \cline{2-10}
                               & 0.7 mA & 1.54 & 1.10 & 1.41 & 0.95 & 1.20 & 1.24 & 1.27 & 1.82 \\ 
        \cline{2-10}
                               & 0.35 mA & 2.03 & 1.40 & 1.47 & 0.95 & 1.21 & 1.18 & 1.17 & 1.60 \\ 
        \specialrule{1.5pt}{0pt}{0pt} 
        \multirow{3}{*}{30 Pa} & 1 mA & 1.68 & 1.18 & 1.33 & 0.99 & 1.22 & 1.41 & 1.22 & 2.10 \\ 
        \cline{2-10}
                               & 0.7 mA & 1.50 & 1.04 & 1.25 & 0.94 & 1.28 & 1.28 & 1.11 & 2.31 \\ 
        \cline{2-10}
                               & 0.35 mA & 1.67 & 0.88 & 1.34 & 1.07 & 1.10 & 1.23 & 1.19 & 1.60 \\ 
        \hline
    \end{tabular}
    \caption{Values $\alpha$ from MSD, $q_p$ from displacement histogram, and $q_v$ from velocity histogram for each of the nine sets of pressure-current conditions.}
    \label{tb:fitparams}
\end{table}

\subsection{Fits to MSD Plots}

\begin{figure}[H]
    \centering
    \includegraphics[width=120mm,height=70mm]{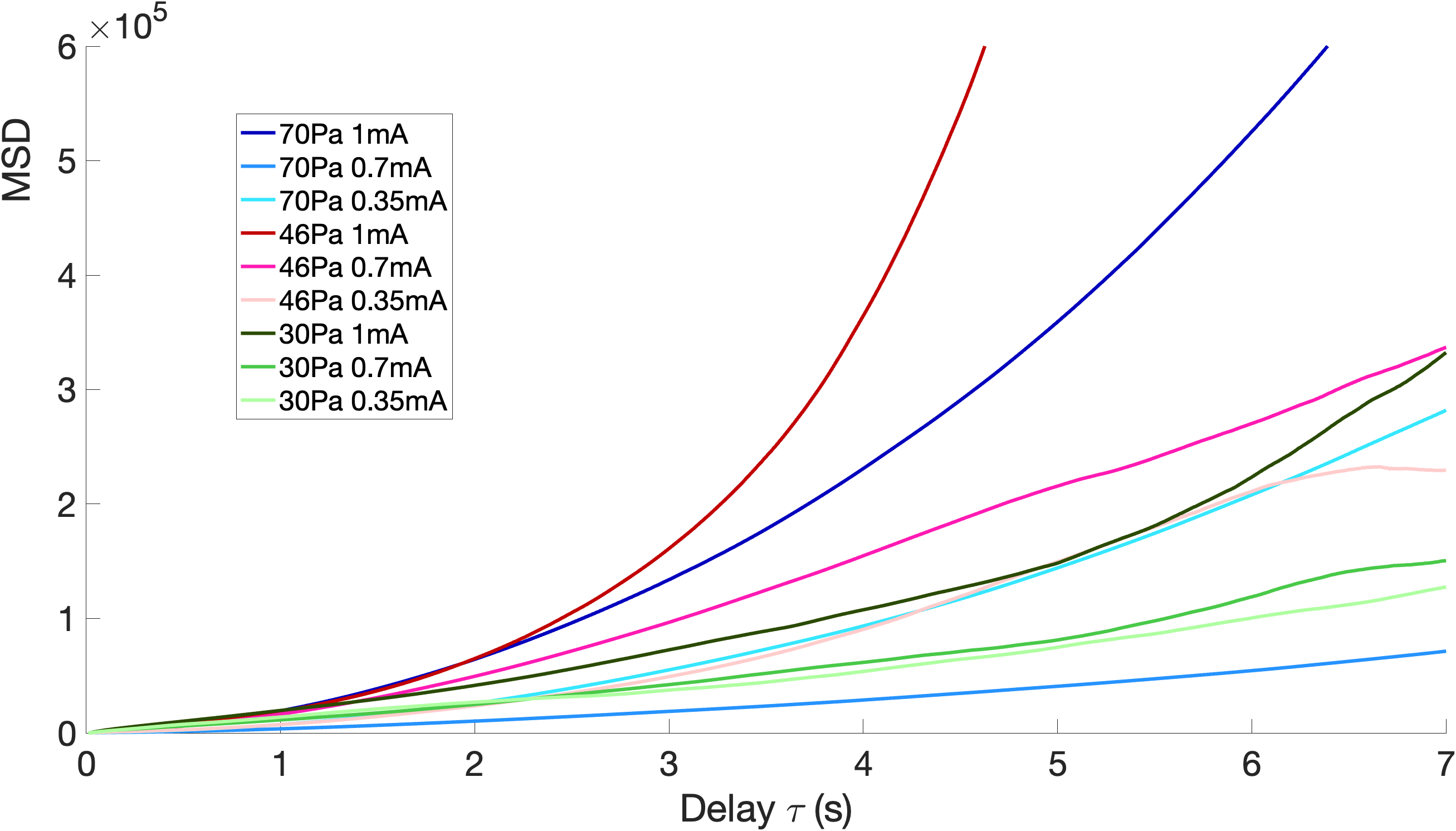}
    \caption{MSD plots for all cases after drift subtraction zoomed in to help distinguish different curves. The y-axis is the mean squared displacement ($\mu m/s$) and the x-axis is time delay in seconds.}
    \label{fig:msd}
\end{figure}

Figure \ref{fig:msd} shows the MSD plots obtained from dust positions within a 2D plane for all cases after drift subtraction. A nonhomogeneous drift subtraction was necessary for the two of the cases: (70 Pa, 0.35 mA) and (46 Pa, 1 mA). For all three pressure cases, it seems like highest current (1 mA) leads to the largest slope of the MSD plots, suggesting an enhanced superdiffusive behavior (see dark red, dark blue, and dark green line on the plot). Normally, in the presence of linear drift in the dust motion, the MSD plot will resemblesuperdiffusivity. An increased dc current is expected to result in an increased electric field strength which would enhance the drift of all charged particles. However, in this case, the fast switching of the dc polarity should still prevent dust response. In addition, examination of the velocity autocorrelation suggests that drift has been successfully subtracted from the particles trajectories, leaving only random motion. Thus, it is likely that the enganced superdiffusive trend is related to the effect that dc current has on the ion stream velocity, charge on the dust, and the resulting ion wakefield structure surrounding each dust particle. 

\vspace{3mm}

\begin{figure}[H]
    \centering
    \begin{subfigure}
        \centering
        \includegraphics[width=70mm,height=40mm]{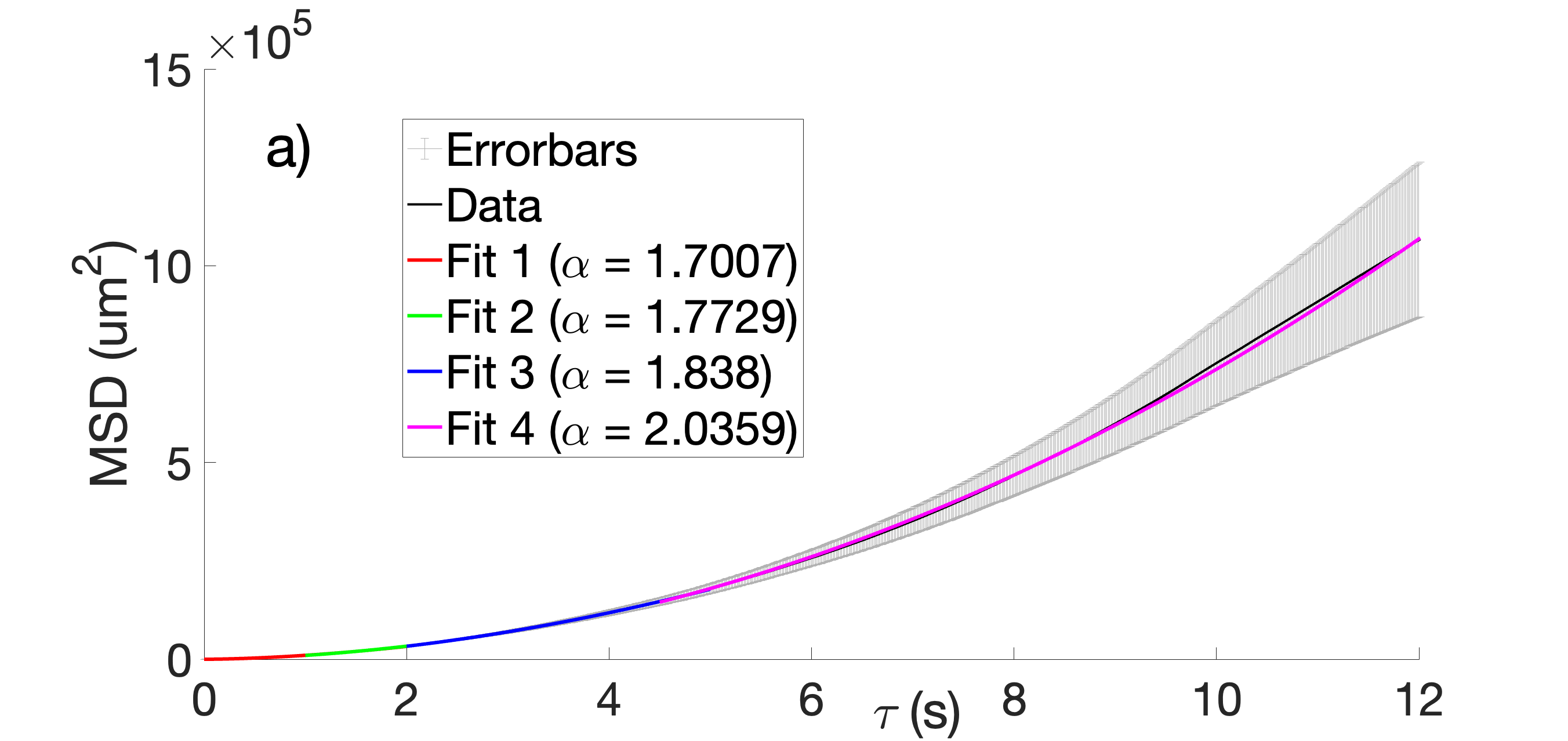}
        \label{fig:xalpha}
    \end{subfigure}
    \begin{subfigure}
        \centering
        \includegraphics[width=70mm,height=40mm]{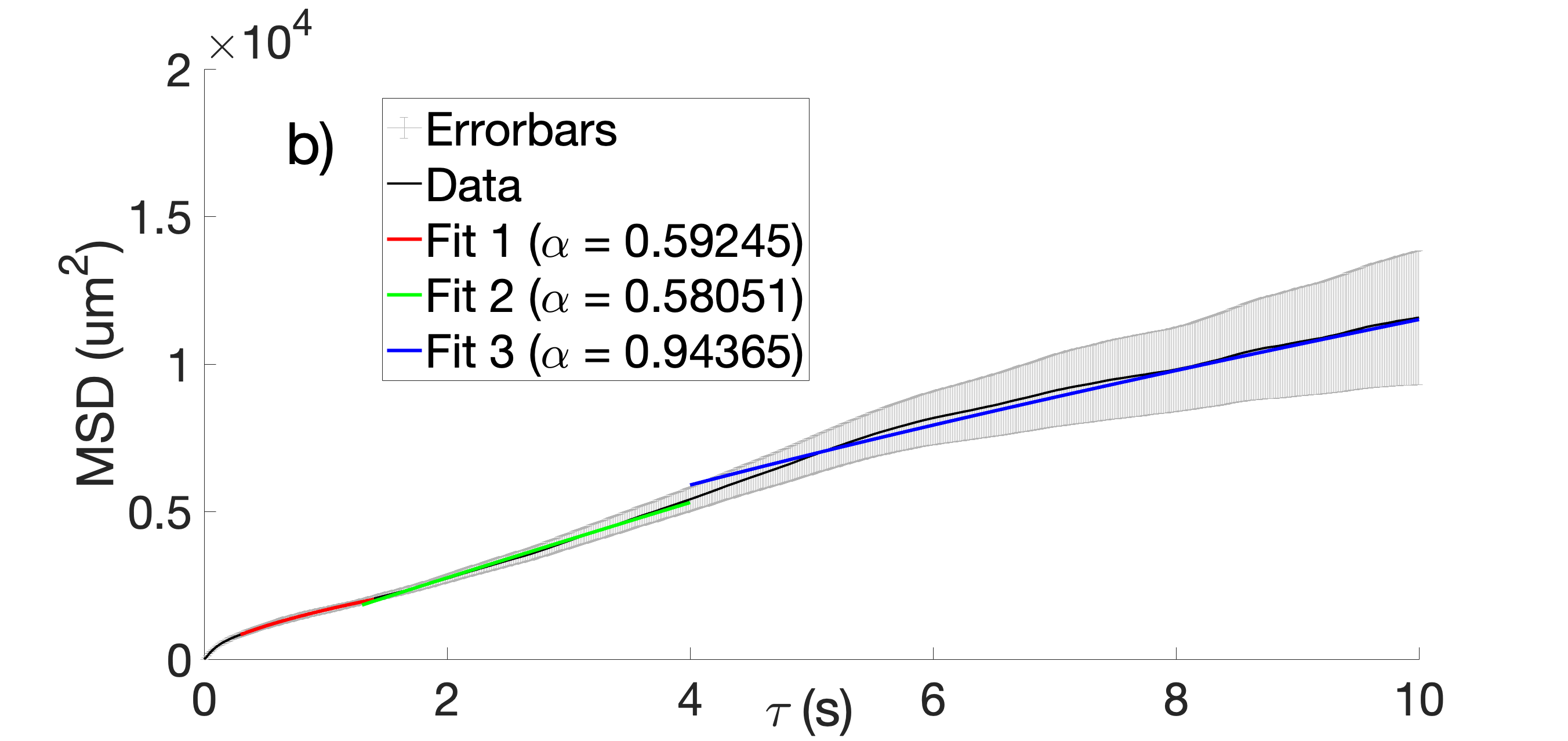}
        \label{fig:plot2}
    \end{subfigure}
    \caption{Fits to mean square displacement for different time delays for the 70Pa, 1mA case. a) MSD plot representative of the axial motion and b) MSD plot representative of the radial motion. The scale of the y axis differ by an order of magnitude between a) and b). This s that more the 2D diffusion is dominated by the axial direction.}
    \label{fig:xymsdcomp}
\end{figure}

To better understand the role of directional anisotropy, we calculated MSDs from particle displacements along $\parallel$ and across $\perp$ the direction of the external electric field. Separate fits were performed at different time delays to assess the role of distinct physical processes. Only data corresponding to time delays smaller than $10s$ was used for the fits. At larger time delays, the standard deviation from the mean MSD increases considerably due to decreasing number of data points. Figure \ref{fig:msd} shows representative plots of $MSD_{\parallel}$ and $MSD_{\perp}$ for the 70Pa, 1mA case. For all pressure-current conditions the $MSD_\parallel$ has a magnitude much greater than $MSD_\perp$ and the plots of $MSD_\parallel$ look almost identical to those of the combined MSDs in Figure \ref{fig:msd}. The $MSD_\parallel$ for all conditions has a positive concavity at all time delays, suggesting superdiffusion. The one exception is the (30 Pa, 0.35 mA) case, which has a brief negative concavity at its start. This dataset had the most noisy trajectories and the least number of data points at long time delays. Thus, the deviation in observed behavior may be due to the poorer quality of the data. The  $MSD_\perp$ for all cases exhibits a brief negative concavity for a delay period of about two seconds, followed by a positive concavity at lager time scales. This suggests that there may be a trapping mechanism causing sub-diffusive behavior at time scales smaller than $2s$. These effects are showcased in the plots in Fig \ref{fig:xymsdcomp}.

Figure \ref{fig:xymsdcomp} is representative of the MSD curves in the parallel and perpendicular directions across all the pressure-current cases. Key features to notice are the larger y-axis scale for the parallel case, the distinct superdiffusive $\alpha>1$ curve for the parallel direction, and subdiffusive $0.1<\tau<~1$ to linear shape ($\tau >5$) of the perpendicular direction. These are consistent across all pressure-current cases.

\begin{figure}[H]
    \centering
        \centering
        \includegraphics[width=150mm,height=55mm]{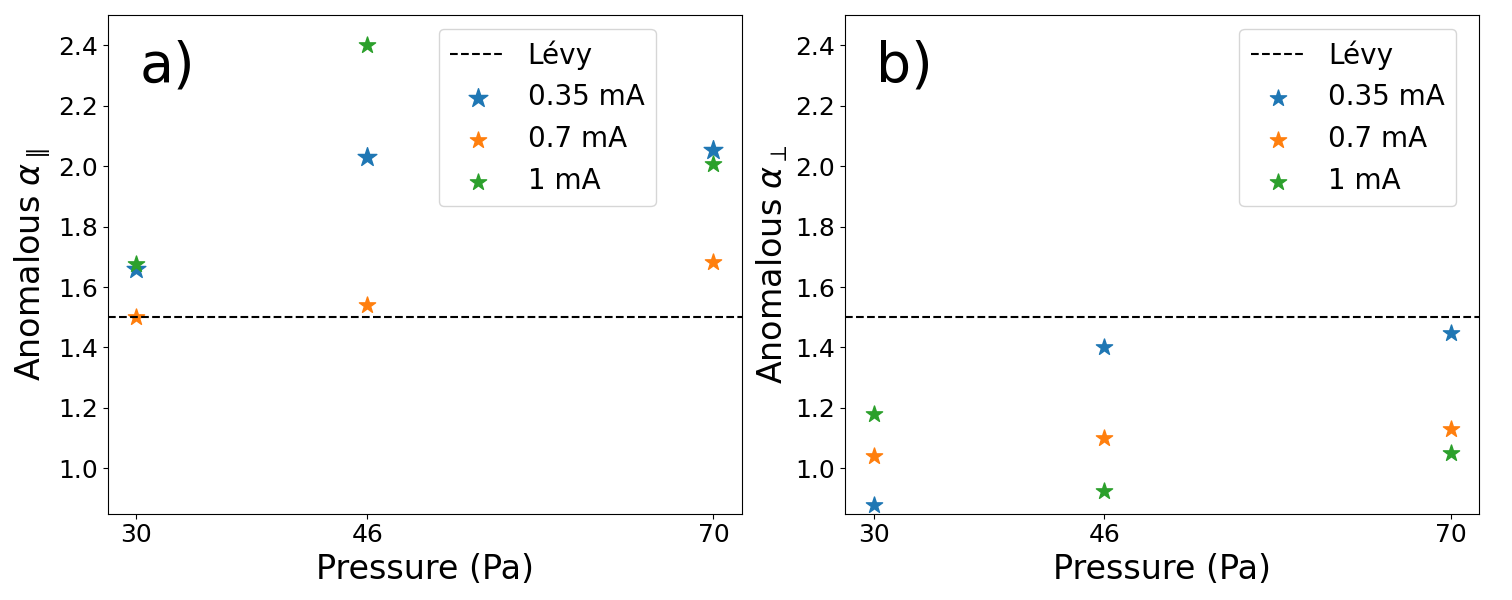}
    \caption{a) $\alpha_{\parallel}$ and b) $\alpha_{\perp}$ for all pressure-current cases.}
    \label{fig:yalpha}
    \label{fig:msdalpha}
\end{figure}

The exponent $\alpha$ extracted from $MSD_{\parallel}$ and $MSD_{\perp}$  at time delay between $4~s$ and $5~s$ is plotted for all cases in Figure \ref{fig:msdalpha}. This time period was chosen as it has the most consistent slope fitting while not being too short of a time delay. Recall that Tsallis’ theory claims that the diffusion is a L\'{e}vy process when $q>5/3$. Using the scaling relation  $MSD=\langle r^2 \rangle \propto \tau^{\frac{2}{3-q_p}}$ \cite{tsallis_introduction_2009} yields the criterion $\alpha>3/2$ for a L\'{e}vy process. Note that this scaling is valid only for $q_p$ obtained from the displacement distributions. A dashed line on Figure \ref{fig:msdalpha} marks the location of $\alpha=3/2$, indicating that almost all distributions for the parallel cases are a L\'{e}vy processes, while the perpendicular direction are not. Later we compare this value of $\alpha$ to that found from $q_p$.

\subsection{Fits to Displacement Histograms}

\begin{figure}[H]
    \centering
    \begin{subfigure}
        \centering
        \includegraphics[width=75mm, height=50mm]{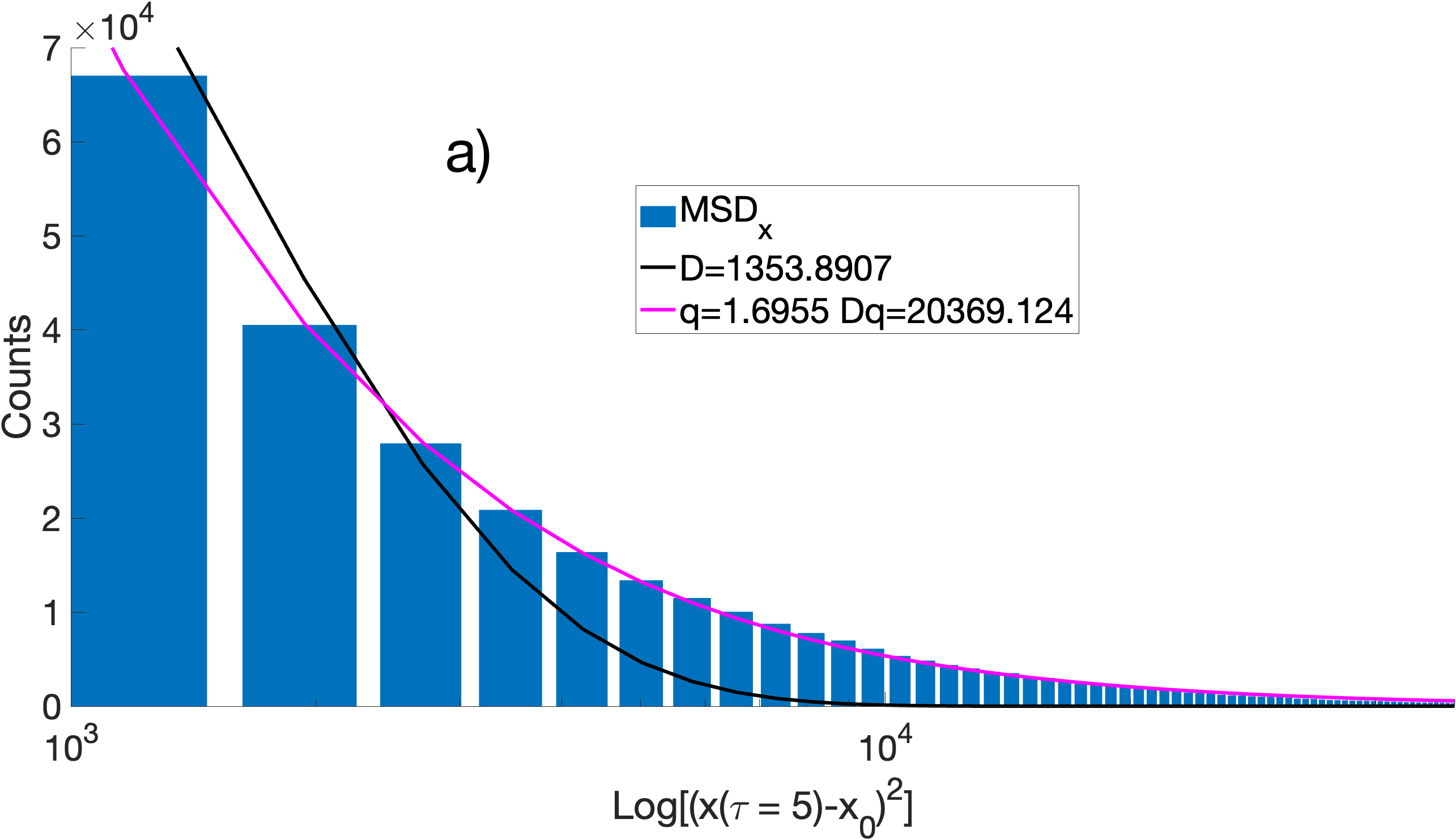}
        \label{fig:plot1}
    \end{subfigure}
    \begin{subfigure}
        \centering
        \includegraphics[width=75mm, height=50mm]{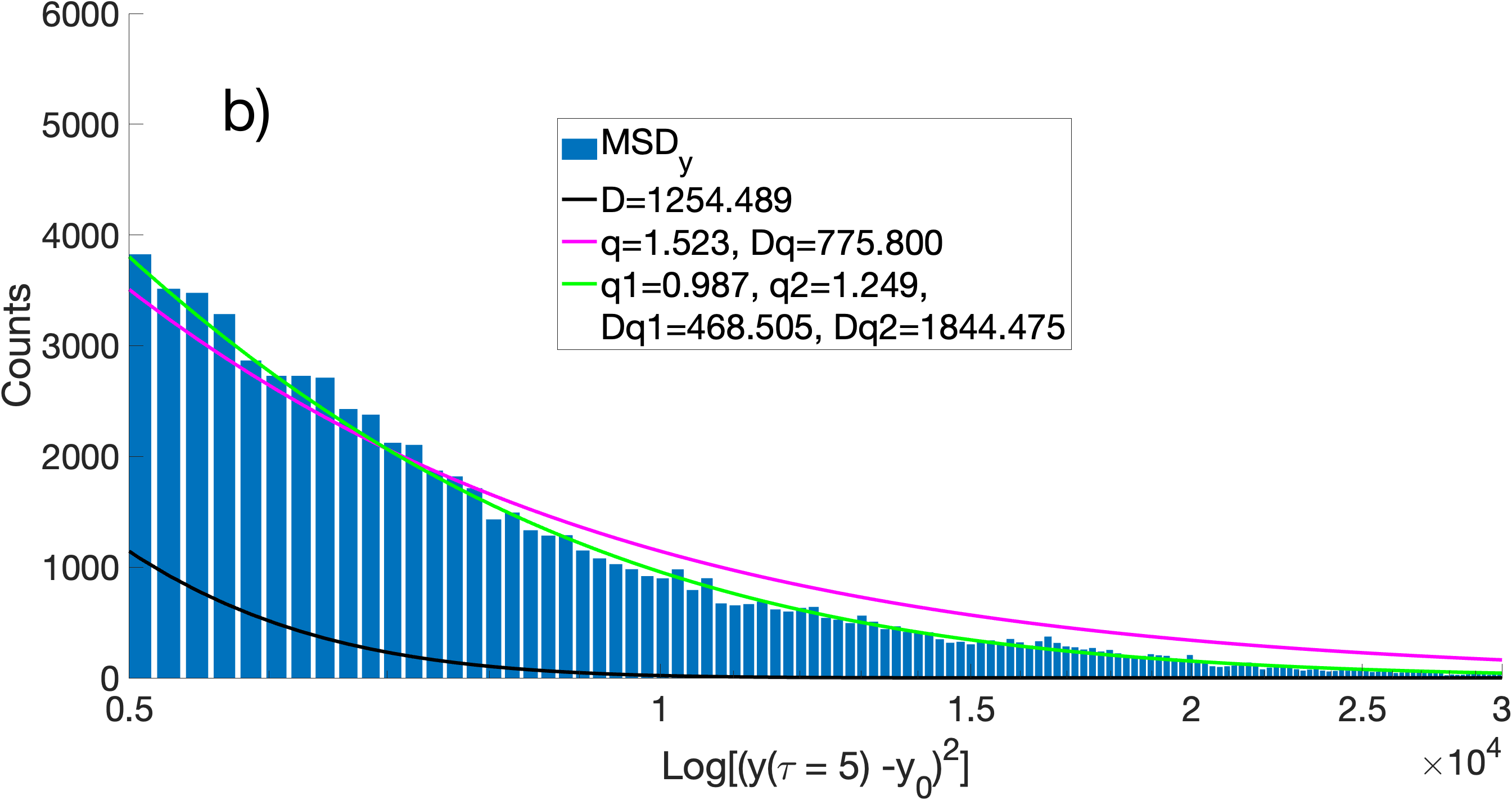}
        \label{fig:plot2}
    \end{subfigure}
    \caption{Position distribution fit example from case 70 Pa 0.7 mA for displacements in a) parallel direction and b) cross field direction plotted in logarithmic scale, shifted horizontally, and zoomed in to accentuate the difference in tail behavior. The data for the position distributions is the positive squared value of $(r(\tau)-r_0)^2$ instead of the velocity distributions $v_r$. Since the data of $(r(\tau)-r_0)^2$ is already squared we fit it to a q-exponential instead of a q-Gaussian.}
    \label{fig:qmsd}
\end{figure}

\begin{figure}[H]
    \centering
        \centering
        \includegraphics[width=160mm,height=90mm]{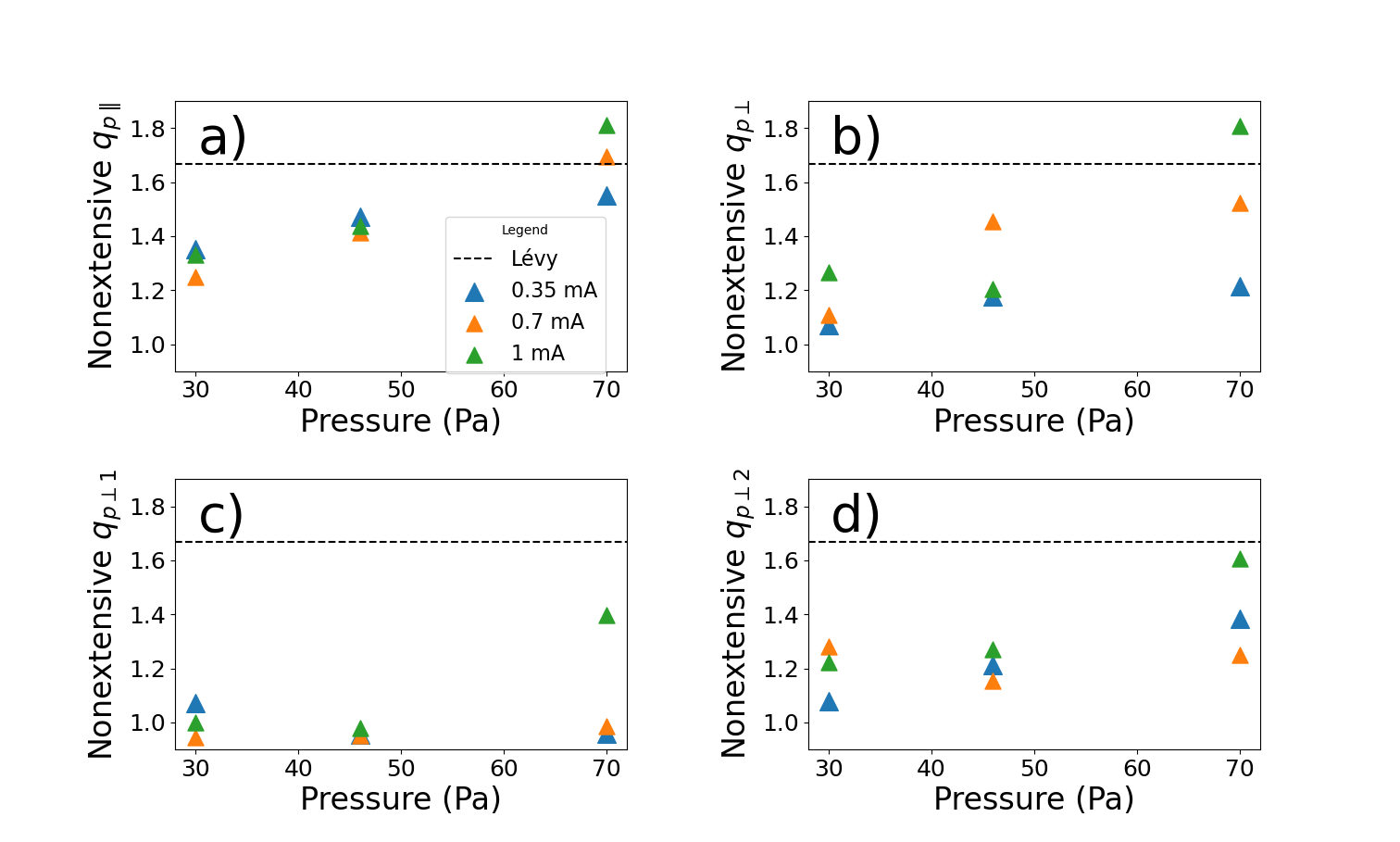}
    \caption{Coefficients extracted from nonextensive $q_p$ fits : a) parallel $q_{p\parallel}$, b) perpendicular single q-Gaussian $q_{p\perp}$, c) perpendicular from Bi-q-Gaussian with Gaussian-like sub-population $q_{p\perp1}$, and d) perpendicular 'tail-halo' $q_{p\perp2}$.}
    \label{fig:qp}
\end{figure}

Representative histograms of the displacements are shown in Figure \ref{fig:qp} for 70 Pa 0.7 mA case. The histograms of displacements parallel to the electric field are best approximated by a single q-Gaussian distribution, while the histograms of cross-field displacements are best approximated by a Bi-q-Gaussian distribution in most cases. These are displacements of all particles at time delay $\tau=5$. The directional difference of the distributions type implies there is anisotropic diffusion. The nonextensive $q_p$ parameters extracted from the displacement histograms are plotted in Figure \ref{fig:qmsd}.

\begin{figure}[H]
    \centering
        \centering
        \includegraphics[width=140mm,height=80mm]{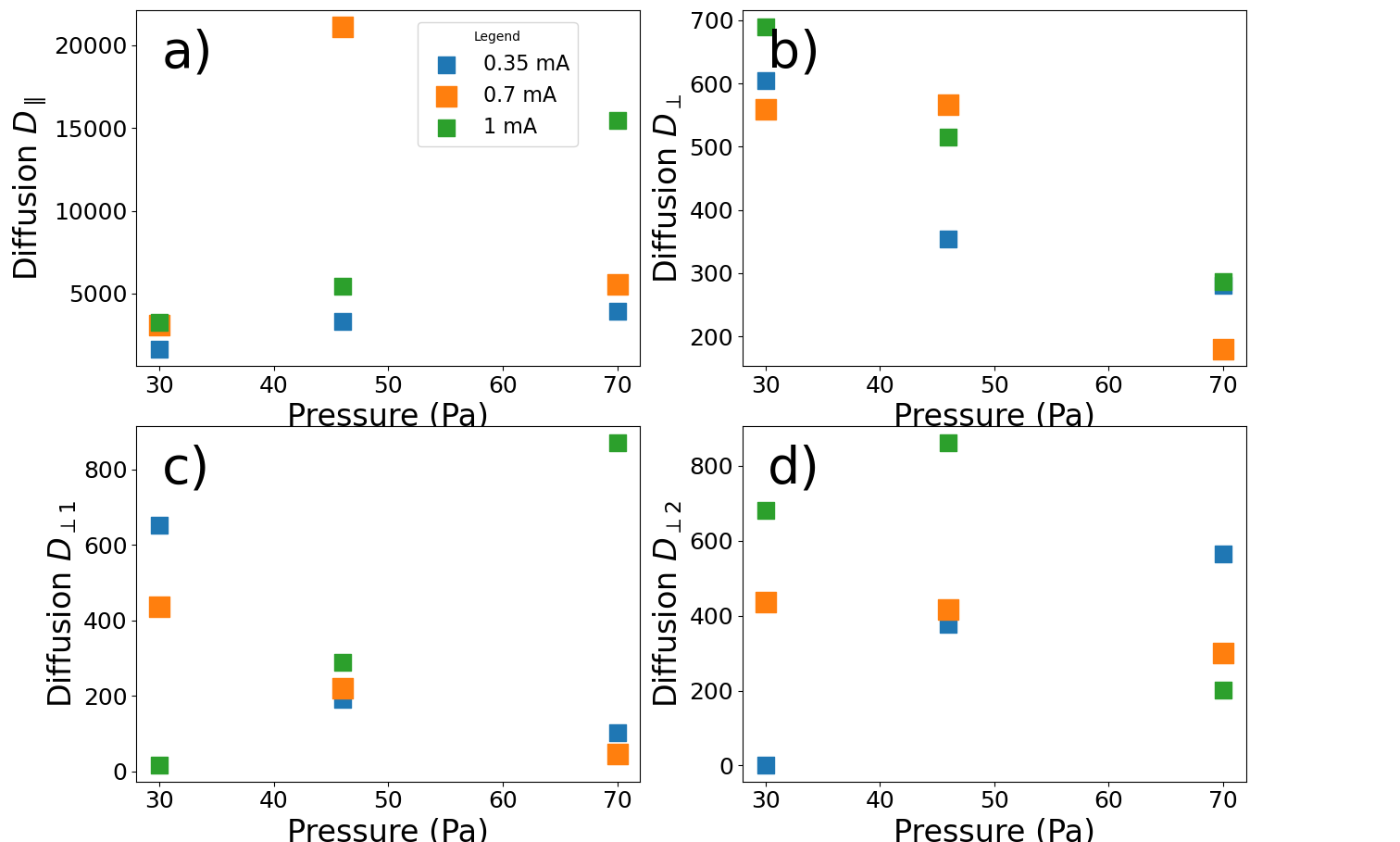}
        \label{fig:plot1}
    \caption{Diffusion coefficients ($/mu m/s$) obtained from fits to the displacement histogram using q-Gaussian and Bi-q-Gaussian. a) parallel, b) perpendicular, c) perpendicular from Gaussian-like sub-population, and d) perpendicular from `tail-halo' population.}
    \label{fig:diffusion}
\end{figure}

Notice that the coefficients for the parallel direction are again larger than for the perpendicular direction. Dashed line indicates a L\'{e}vy process. The coefficients extracted from the histogram of parallel displacements suggest that two of the 70Pa cases are consistent with a L\'{e}vy process. A crossover to a L\'{e}vy process is also seen for the cross-field direction in the 70 Pa, 1 mA case if a single q-Gaussian is fitted to the histogram of cross-field displacements. However, the more accurate Bi-q-Gaussian fit to the cross-field displacement histograms suggest that the process is likely superdiffusive but not L\'{e}vy.  These results are in qualitative agreement with the conclusions drown from the $\alpha$ coefficients extracted from the MSD fits at time delay $\tau=5$ shown in Figure \ref{fig:msdalpha}. It seems that higher pressures create larger spread in the coefficients and exhibit a larger difference between the parallel and perpendicular directions suggesting a more pronounced anisotropic effect. The values for the diffusion $D \left( \frac{\mu m^2}{s} \right)$ calculated from Eq. \ref{eq:qvariance}, are shown in Figure \ref{fig:diffusion}. Here we can see that, in the direction parallel to the field, an increase in pressure causes diffusion to increase. However, in the cross-field direction, the interpretation is less straightforward. $D_{\perp1}$ and $D_{\perp}$ generally decrease with pressure with some exceptions. $D_{\perp2}$ does not show a clear trend with pressure or current.

\begin{figure}[H]
    \centering
        \centering
        \includegraphics[width=140mm,height=80mm]{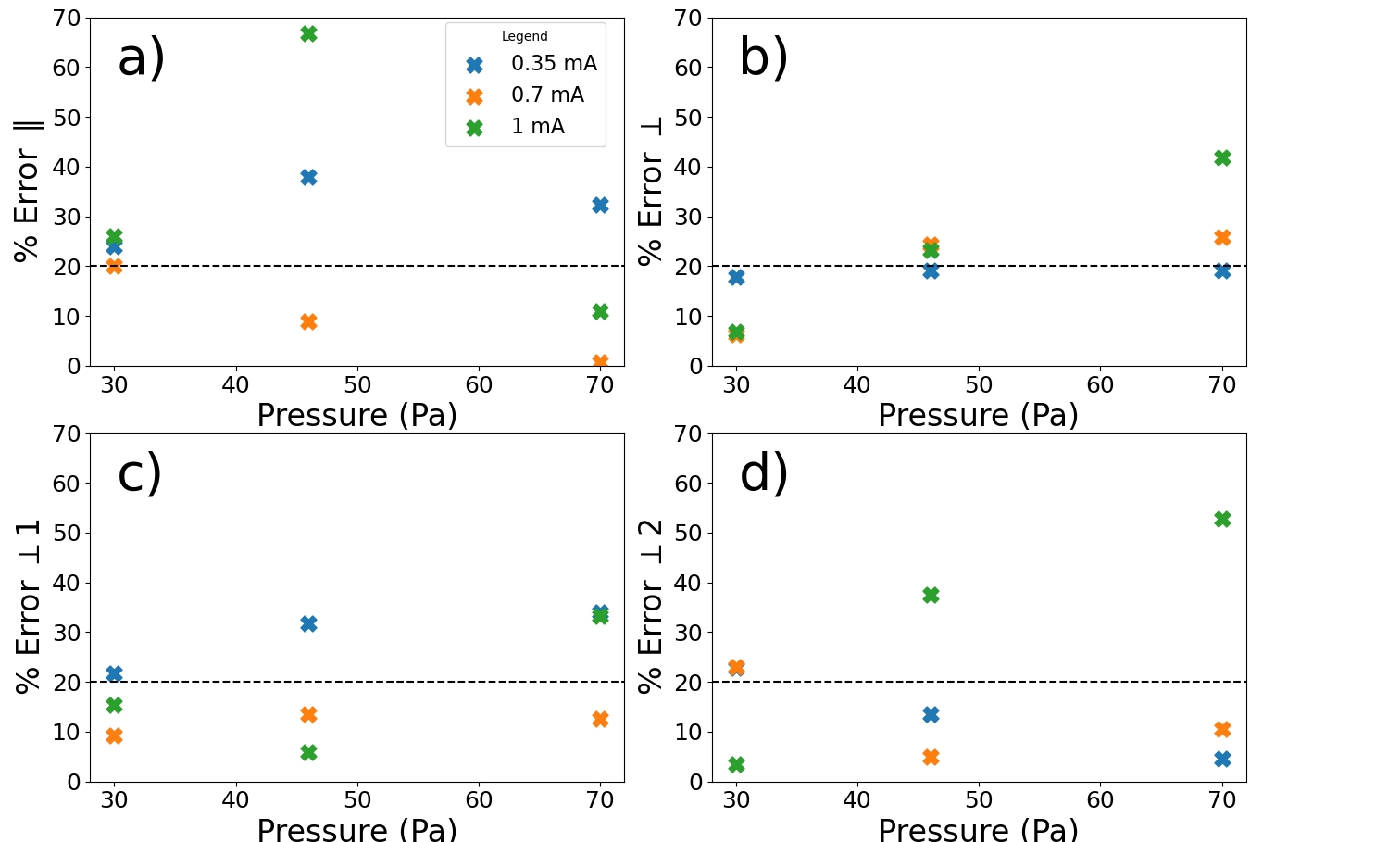}
    \caption{Percent difference between $\alpha$ found from MSD curves and $\alpha$ found from $q_p$ using Eq. \ref{eq:alpha2q}. a) parallel component, b) perpendicular component from a single q-Gaussian, c) Gaussian-like `sub-population', and d) `tail-halo'.  A dashed line at 20\% is used to help compare the figures.}
    \label{fig:q vs alpha error}
\end{figure}

 Next, we examined how the exponent $\alpha$ found directly from MSD fits compares to $\alpha$ calculated from the displacement histogram fits using the scaling in Eq. \ref{eq:alpha2q}. Figure \ref{fig:q vs alpha error} shows the percent difference $\frac{\alpha_{MSD}-\alpha_{q_p}}{\alpha_{q_p}}$. We can see that on average the error is lower, indicated by the dashed line at 20\%, for the cases where a Bi-q-Gaussian fit was used rather than a single q-Gaussian. This seems to be in agreement with findings from \cite{liu_non-gaussian_2008}.

\subsection{Velocity Distribution Plots}

\begin{figure}[H]
    \centering
    \begin{subfigure}
        \centering
        \includegraphics[width=75mm,height=40mm]{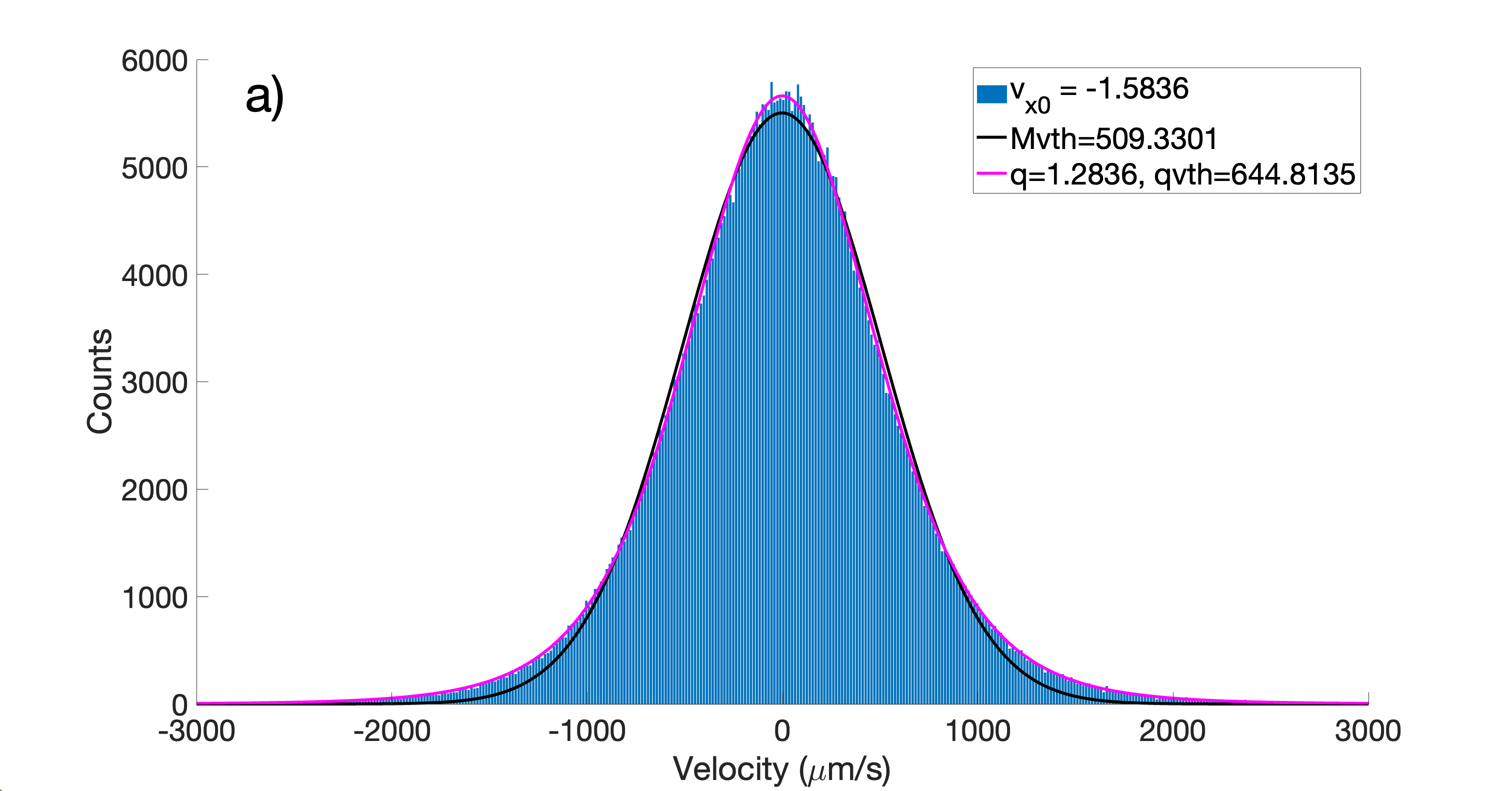}
        \label{fig:plot1}
    \end{subfigure}
    \begin{subfigure}
        \centering
        \includegraphics[width=75mm,height=40mm]{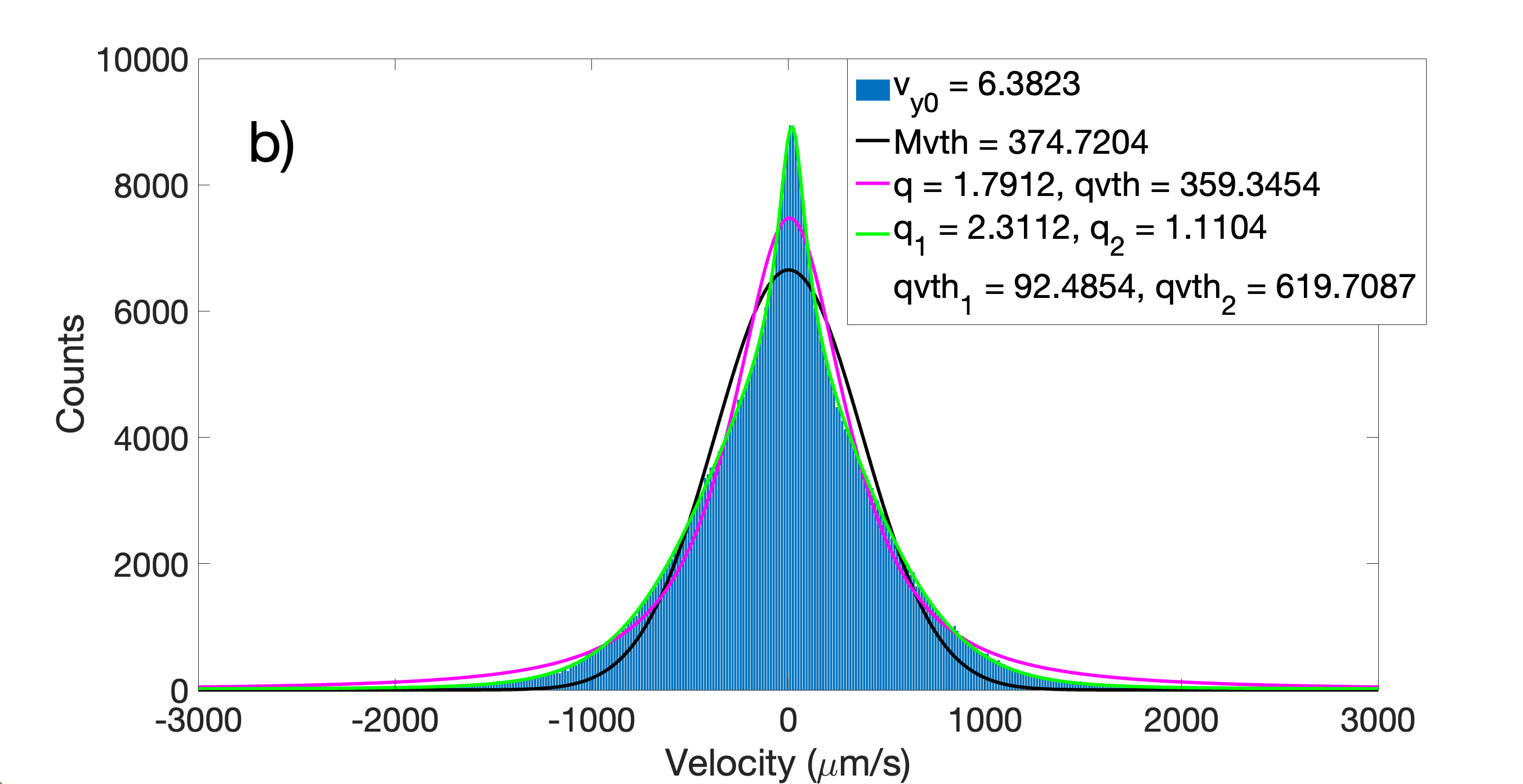}
        \label{fig:plot2}
    \end{subfigure}
    \caption{Velocity component histograms and distribution fits for the 30 Pa, 0.7mA case. a) Velocity components along the direction of the electric field. b) Velocity components in the cross-field direction.}
    \label{fig:vhist}
\end{figure}

Next we examine Figure \ref{fig:vhist}, which shows representative histograms and distribution fits for the velocity components parallel and perpendicular to the direction of the external electric field. These plots show that a Gaussian or a Maxwellian distribution (black line) does not approximate the `fat' tails of the histograms, while a q-Gaussian (magenta line) and a Bi-q-Gaussian (green line) provide good fits to the $\parallel$ and the $\perp$ velocity histograms, respectively. It is again observed that the distribution features change substantially with direction, suggesting anisotropic dust diffusion caused by the external electric field. Similar to the displacement data, we observe that the velocity histograms in the cross-field direction are best approximated by a Bi-q-Gaussian fit, suggesting a superposition of two distinct diffusion processes. These trends in histogram shapes are observed for all pressure-current cases. This will be further discussed in Sec. \ref{subsec:PerpHist}.

\begin{figure}[H]
    \centering
        \centering
        \includegraphics[width=140mm,height=80mm]{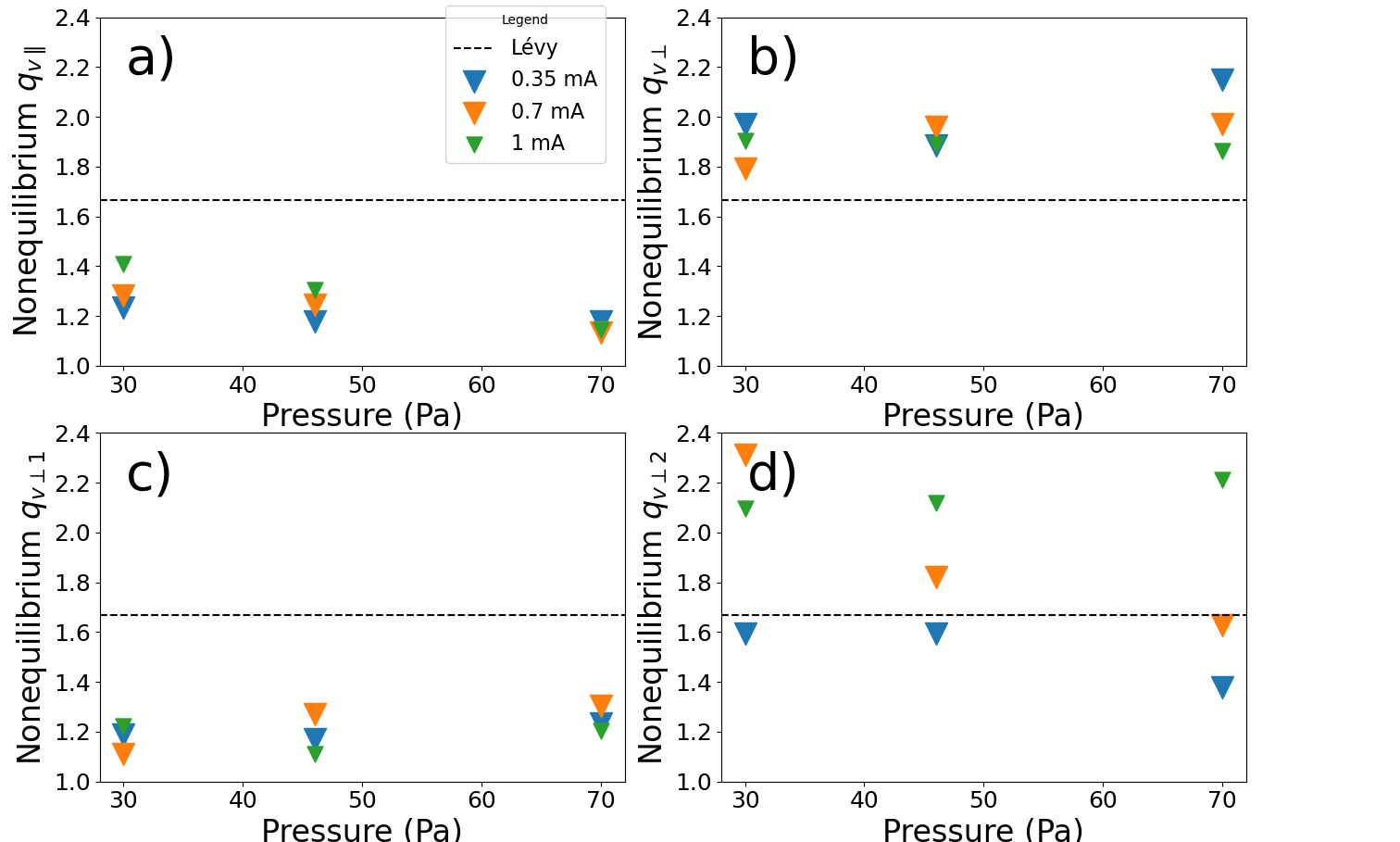}
    \caption{Nonequilibrium coefficients extracted from fits to velocity histograms for all pressure-current cases. a) $q_{v\parallel}$, b) single q-Gaussian $q_{v\perp}$, c) $q_{v\perp1}$ is a Gaussian-like sub-population, and d) $q_{v\perp2}$ is a 'halo-tail'.}
    \label{fig:q_nonequilibrium}
\end{figure}

Figure \ref{fig:q_nonequilibrium} shows the parallel and perpendicular nonextensive q parameters obtained from the velocity histograms for each pressure-current case. A slight negative slope is observed for increasing pressure for $q_{v\parallel}$ and none of the coefficients are above the L\'{e}vy line. The single q-Gaussian fits for the cross field velocity histograms had an $R^2 = 0.973$ and a normalized root mean square error $NRMSE = 0.027$ while the Bi-q-Gaussian fit had $R^2 = 0.993$ and $NRMSE = 0.013$. A larger value of $R^2$ and lower value of $NRMSE$ imply better fit thus the single q-Gaussian fits for the cross field velocity histograms are not as representative of the system as well as the Bi-q-Gaussian. With the Bi-q-Gaussian fits the $q_{\perp1}$ values show little change with plasma conditions (most of them are around 1.2), while $q_\perp2$ values seem to decrease with increasing pressure, except for the 1 mA case. The parameter $q_v$ provides information on the equilibrium of the system, suggesting that in the direction parallel to the electric field, the dust ensemble is close to an equilibrium as seen in Figure\ref{fig:q_nonequilibrium} a), despite the strong superdiffusion observed in Figure \ref{fig:qp} a).The trends in the cross-field direction are opposite, suggesting that the identification of the two distinct  diffusion patterns drive the system away from equilibrium. This will be further discussed in Sec. \ref{subsec:vdfs}. 

\begin{figure}[H]
    \centering
    \centering
        \centering
        \includegraphics[width=130mm,height=50mm]{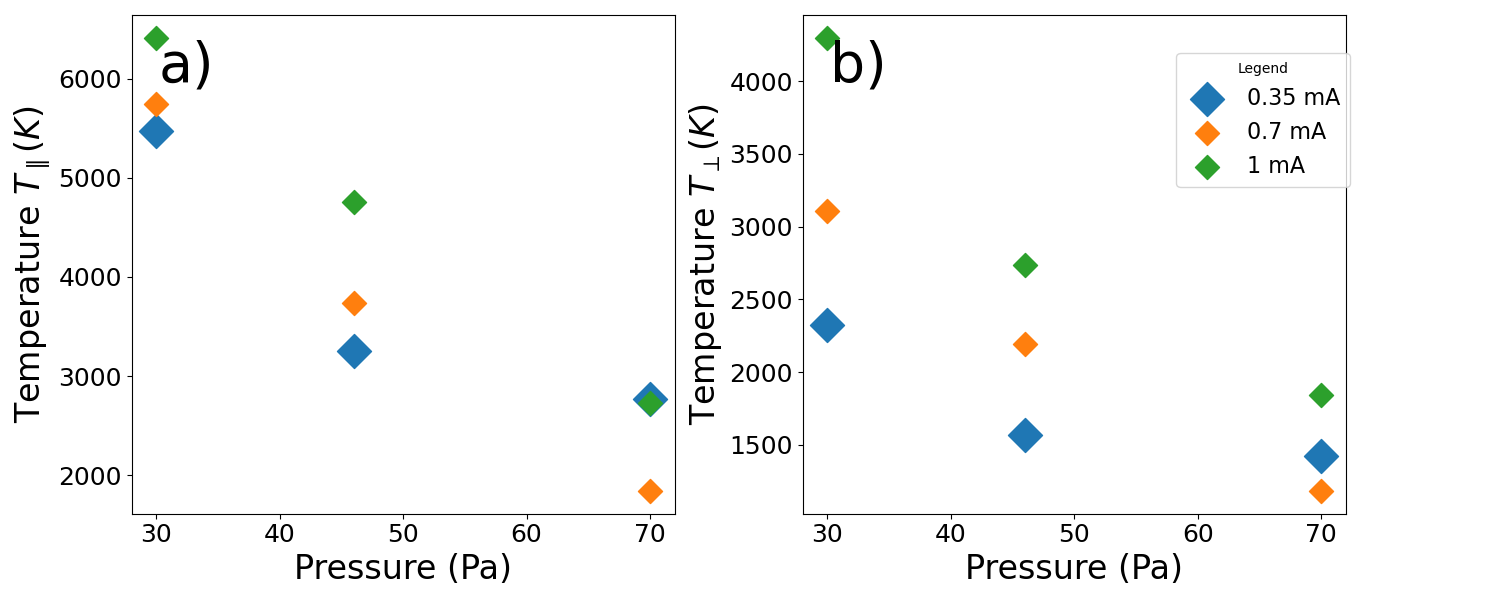}
    \caption{Temperatures for all pressure-current cases found using a Maxwellian fit to the a) parallel and b) perpendicular components of the velocity histograms.}
    \label{fig:vthM}
\end{figure}

\begin{figure}[H]
    \centering
        \centering
        \includegraphics[width=140mm,height=80mm]{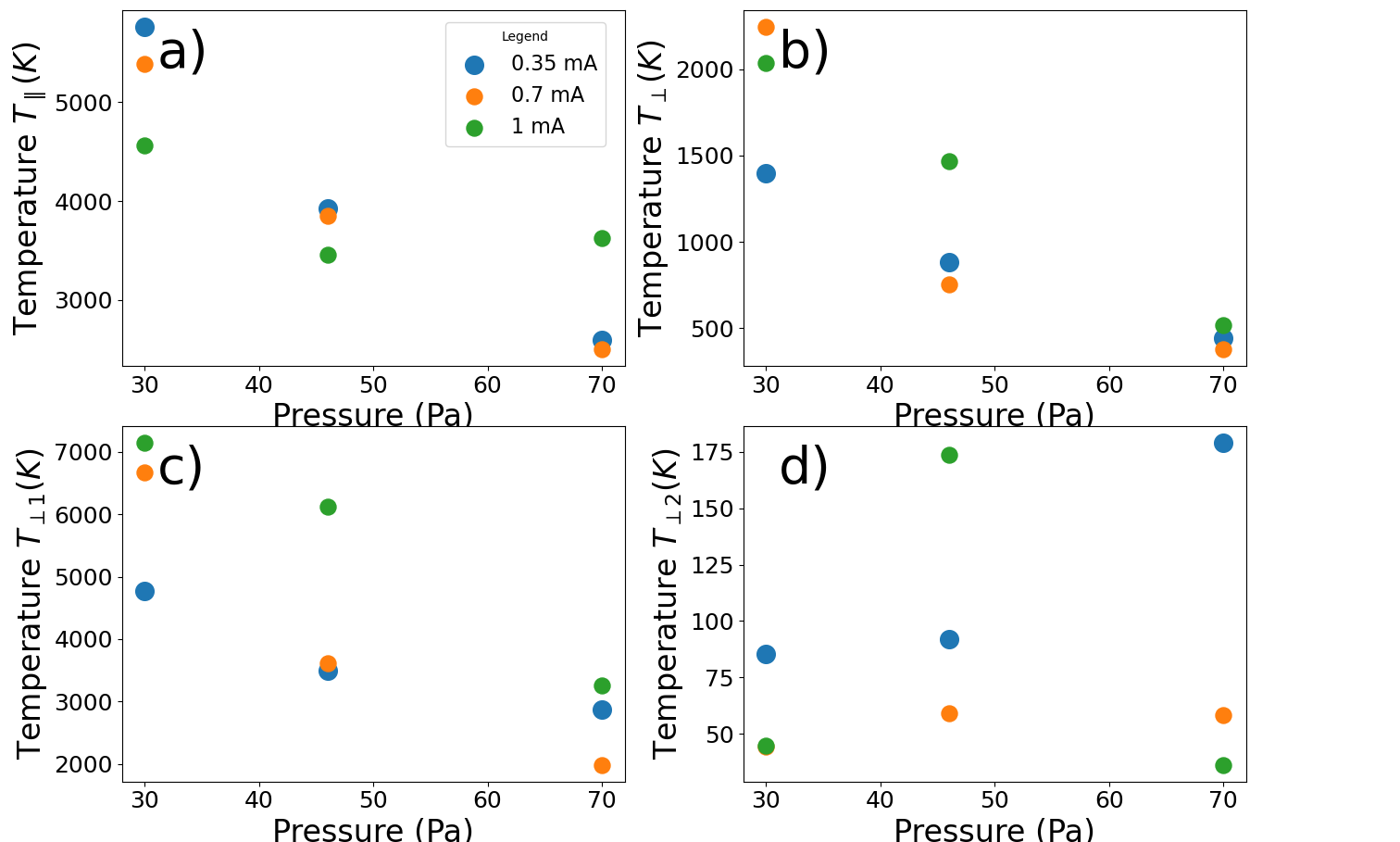}
        \label{fig:plot1}
    \caption{Temperatures $T_{Mq}$ for all pressure-current cases found using non-Maxwellian fits to the parallel and perpendicular components of the velocity histograms. Temperature in the parallel direction a) was obtained using a single q-Gaussian fit. Temperatures in the perpendicular direction was found using a q-Gaussian b) and a Bi-q-Gaussian with a Gaussian-like sub-population fit c) and perpendicular `tail-halo' fit d).}
    \label{fig:Tq}
\end{figure}

 The velocity distribution fits were used to extract dust temperatures for the directions parallel and perpendicular to the electric field. Figures \ref{fig:vthM} and \ref{fig:Tq} shows the temperatures found using a Maxwellian fit, and a q-Gaussian fit, respectively. 

Kinetic temperature is typically defined as the variance of a Maxwellian distribution. Since the velocity histograms for the dust in the PK-4 experiment are best approximated by a non-Maxwellian, the notion of "temperature" here is not well defined. The range of kinetic temperatures found from the Maxwellian fits (Figure \ref{fig:vthM}) is $2000K-6500K$, or about $0.2eV-0.6eV$, suggesting that the majority of the dust particles have very high kinetic energy even though the experiment is at room temperature and the pressure is low. Anomalously high dust temperature is a known phenomena that is caused by electrostatic fluctuations \cite{Avinash2011}. In Figure \ref{fig:Tq}, we used Eq. \ref{eq:qvariance}, $T_q(\frac{5q-3}{2})=T_{Mq}$ (from \cite{Bilal2023}), since it provides a relation between the temperature or thermal velocity found from a q-Gaussian and to a `Maxwellian-like' temperature.

 As shown in Figure \ref{fig:vhist} b), the perpendicular velocity histograms are best approximated by a Bi-q-Gaussian fit, where the the bulk distribution (the one with the larger variance and smaller $q_v$ value) more closely resembles a Gaussian and, thus, can be used to extract the thermal velocity. The second distribution in the Bi-q-Gaussian accounts for the "fat tails". Thus, in Figure \ref{fig:Tq}, $T_{\perp1}$ is extracted from a Gaussian-like sub-population distribution, while $T_{\perp2}$ is obtained from fitting to a 'tail-halo' distribution. It may seem unintuitive why the Gaussian sub-population would have higher temperature values, while the 'halo-tails' have lower temperature values. This will be discussed in \ref{subsec:PerpHist}. Essentially we convert a variance from a distribution which has some 'tailed-ness' to a variance which does not have any, so that we can use the same definition of temperature. We argue that this is not a different treatment of temperature in dusty plasma, but an attempt at calculating temperature more carefully with nonextensive statistics. This allows us to derive more accurate dust temperatures for the PK-4 dusty plasma with values in the range $2500K-5500K$, or about $0.2eV-0.5eV$, which is smaller than the typically calculated values of $~10-300eV$ \cite{Avinash2011}. 

\vspace{3mm}

Finally, for velocity q-distributions, Jiulin and Haining \cite{du_nonextensivity_2004,haining_nonextensive_2014} have derived the following expression for $q_v$ in terms of a thermophoretic force and Lorentz 
force 

\begin{equation}
    q_v = 1 - \frac{k_B}{e}\frac{\nabla T_s}{(\nabla \phi + \vec{v} \times \vec{B})}.
    \label{eq:duqv}
\end{equation}

$T_s$ is the temperature for each species: electron, ion, or dust. In the PK-4, there is no magnetic field, thus, $\vec{v} \times \vec{B} = 0$. In the absence of a temperature gradient, $q_v=1$ and the system is in equilibrium, i.e., the velocity histograms should be best described by a Maxwell-Boltzmann distribution. Equation \ref{eq:duqv} has has not yet been verified against any experimental measurements, though we will give a qualitative argument for its possible credibility.

\subsection{Domain Velocity Distribution Plots}

\begin{figure}[ht]
    \centering
    \begin{subfigure}
        \centering
        \includegraphics[width=75mm,height=45mm]{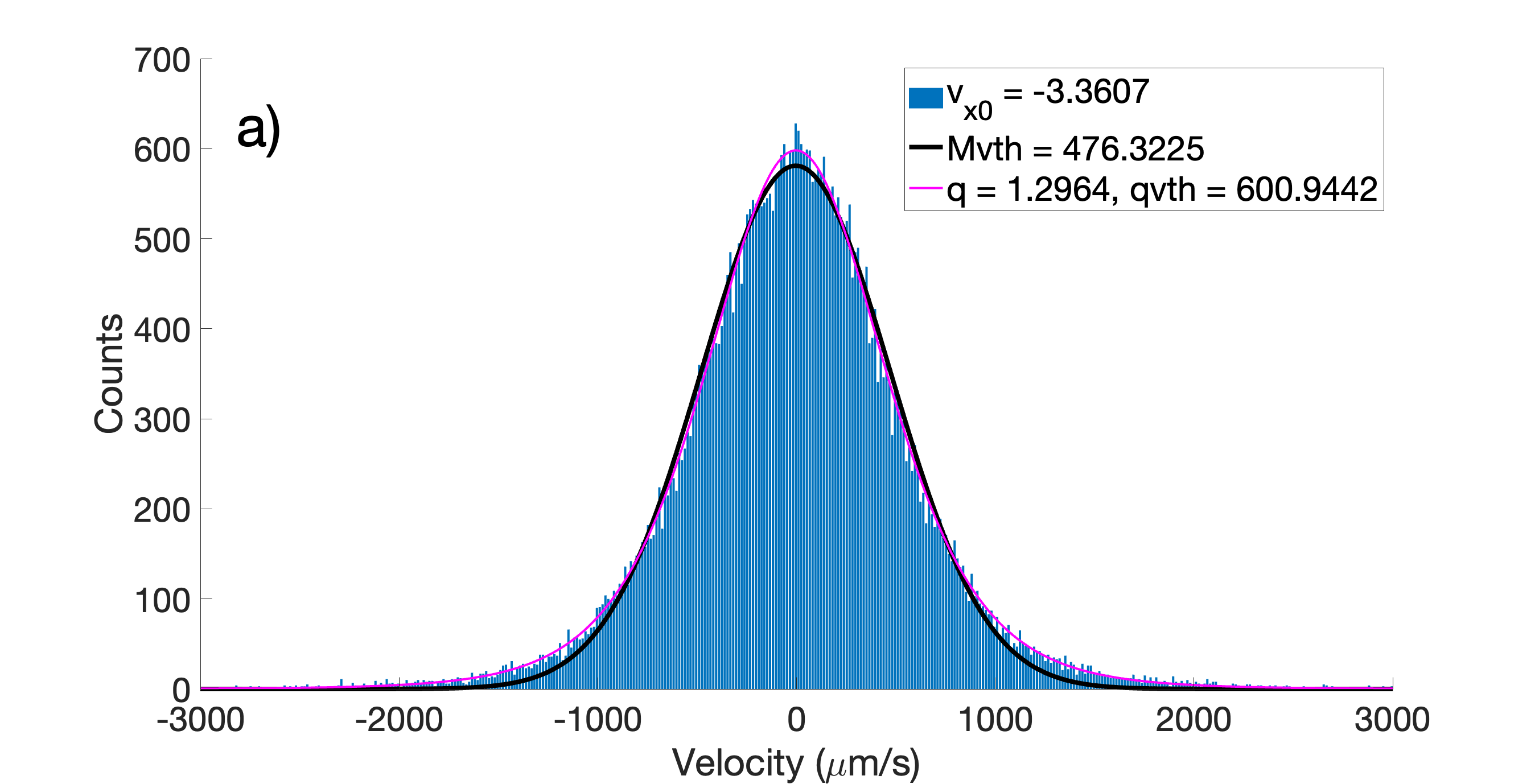}
        \label{fig:plot1}
    \end{subfigure}
    \begin{subfigure}
        \centering
        \includegraphics[width=75mm,height=45mm]{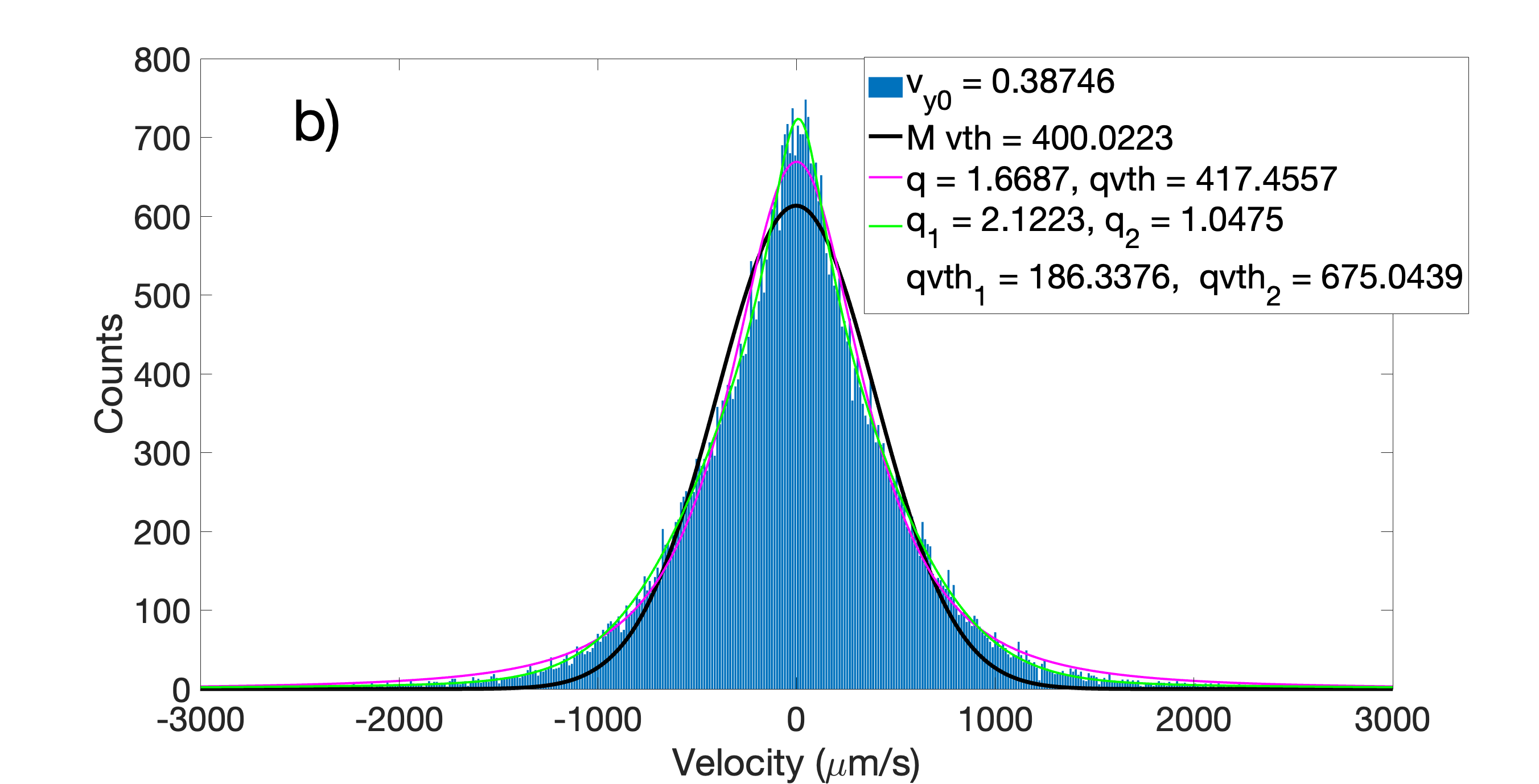}
        \label{fig:plot2}
    \end{subfigure}
    \caption{Velocity distributions for a) $\parallel$ and b) $\perp$ directions obtained from  histogram fits of domain 11 for the case 30 Pa 0.7 mA. Notice that the q-Gaussian (magenta) curve provides a much closer fit than the one shown in Figure \ref{fig:vhist}.}
    \label{fig:domain velocity}
\end{figure}

Here we provide physical arguments why the Bi-q-Gaussian fit for the perpendicular histograms is needed. First, we consider the domain separation on Figure \ref{fig:nhds} A) which was made for the NHDS method and reconstruct the velocity histogram in each domain. Figure \ref{fig:domain velocity} shows the velocity distribution for domain 11 (dark blue in Figure \ref{fig:nhds} A) in both the parallel and perpendicular directions for from the 30 Pa 0.7 mA case.

Although the Maxwellian fit still underpredicts the tails of the distribution, the single q-Gaussian fits the perpendicular histogram data much more closely than in Figure \ref{fig:vhist}, though the Bi-q-Gaussian is also a reasonable fit. This phenomenon is apparent in all of the other domains and for all pressure cases with current $0.7mA$. Other current cases were not analyzed. We attribute these differences to the transition between local and global dynamics as the system size increases. The nonextensive Tsallis parameter can be used to assess the nonequilibrium state of the system with equilibrium represented by $q=1$ and nonequilibrium by $q\neq 1$. The different local domains in a system may be more or less in equilibrium depending on the properties of the corresponding local distribution functions. As the sample of velocities used to create the histogram is increased to include multiple domains with varying equilibrium properties, the conclusions about global diffusion and thermodynamics may be significantly altered by the averaging process. A smaller domain size is more likely to provide accurate information on the nonextensive parameter $q_v$, whether the particles in the small region are in equilibrium or not. Of course, this requires that the number of data points in the smaller domain is sufficient to yield statistically significant results. Table \ref{table:domains} provides the values of all the nonequilibrium $q_v$ parameters in each domain with each table entry representing the domain location in the field of view of the particle tracks, (Figure \ref{fig:nhds} a). This yields a spatial map of the equilibrium properties at different pressure of the PK-4 dust cloud. Table \ref{table:domains} shows that regions near the outer part of the dust cloud are typically further away from equilibrium than regions within the central region.

\newpage

\begin{center}
  {\Large{Nonequilibrium $q_{v\perp}$} in each domain}  
\end{center}

\begin{table}[H]
  \captionsetup{labelsep=none} 
  \begin{minipage}{0.3\textwidth}
    \centering
    \begin{tabular}{|c|c|c|c|}
      \hline
      1.57 & 1.63 & 1.47 & 1.45 \\
      \hline
      1.51 & 1.51 & 1.45 & 1.44 \\
      \hline
      1.66 & 1.71 & 1.79 & 1.55 \\
      \hline
    \end{tabular}
    \caption*{30 Pa 0.7 mA}
  \end{minipage}
  \hfill
  \begin{minipage}{0.3\textwidth}
    \centering
    \begin{tabular}{|c|c|c|c|}
      \hline
      1.98 & 2.05 & 2.01 & 1.83 \\
      \hline
      1.63 & 1.68 & 1.68 & 1.58 \\
      \hline
      1.63 & 1.66 & 1.62 & 1.60 \\
      \hline
    \end{tabular}
    \caption*{46 Pa 0.7 mA}
  \end{minipage}
  \hfill
  \begin{minipage}{0.3\textwidth}
    \centering
    \begin{tabular}{|c|c|c|c|}
      \hline
      1.57 & 1.63 & 1.47 & 1.45 \\
      \hline
      1.51 & 1.51 & 1.45 & 1.44 \\
      \hline
      1.66 & 1.71 & 1.79 & 1.55 \\
      \hline
    \end{tabular}
    \caption*{70 Pa 0.7 mA}
  \end{minipage}
  
  \captionsetup{labelsep=colon} 
  \caption{The three tables show $q_v$ placed in the table at the corresponding domain showing a spatial map of the equilibrium.}
  \label{table:domains}
\end{table}

\vspace{5mm}

\begin{figure}[H]
    \centering
    \begin{subfigure}
        \centering        \includegraphics[width=60mm,height=40mm]{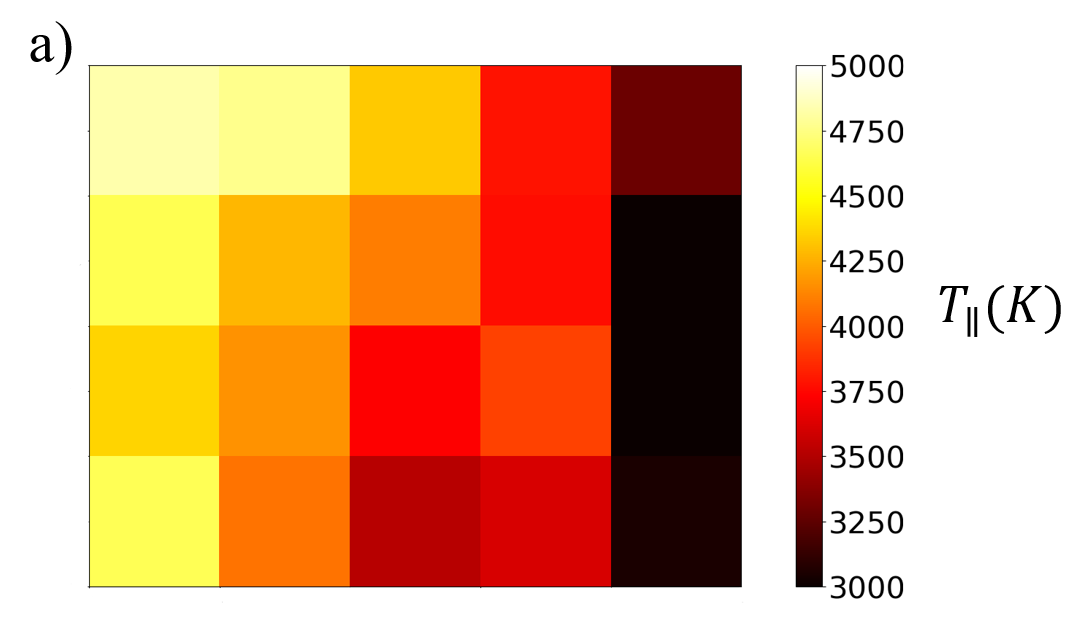}
        \label{fig:T perp gradient}
    \end{subfigure}
    \begin{subfigure}
        \centering        \includegraphics[width=60mm,height=40mm]{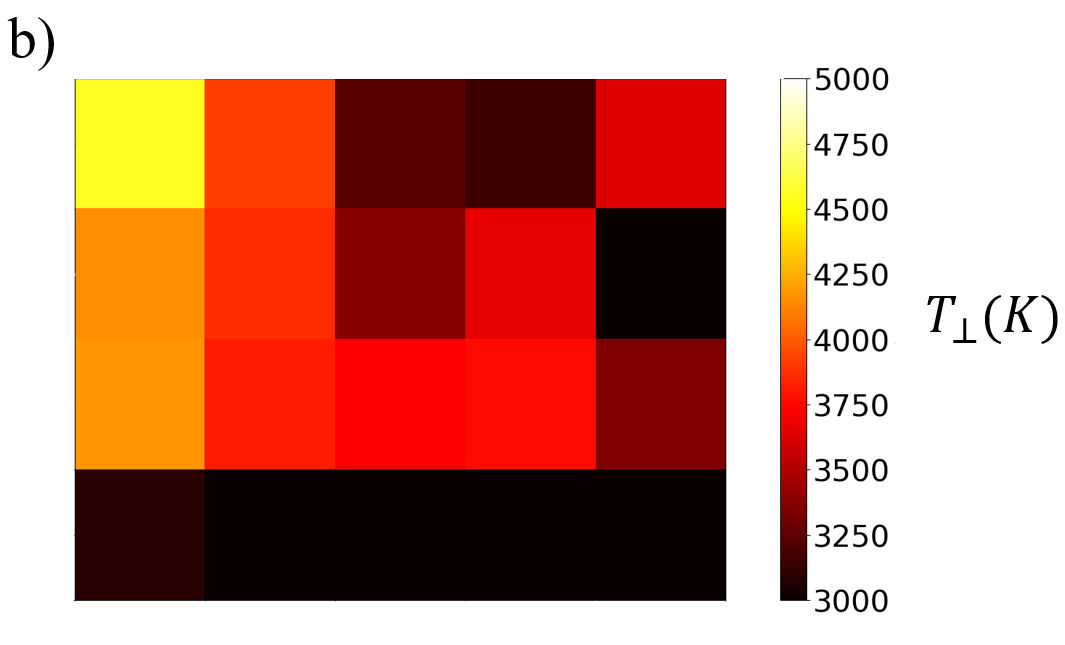}
        \label{fig:T para gradient}
    \end{subfigure}
    \caption{a) $T_\|$ and b) $T_\perp$ calculated in each domain showing temperature gradients (30 Pa 0.7 mA).}
    \label{fig:temp gradient domains}
\end{figure}

\begin{figure}[H] 
    \centering
    \includegraphics[width=80mm]{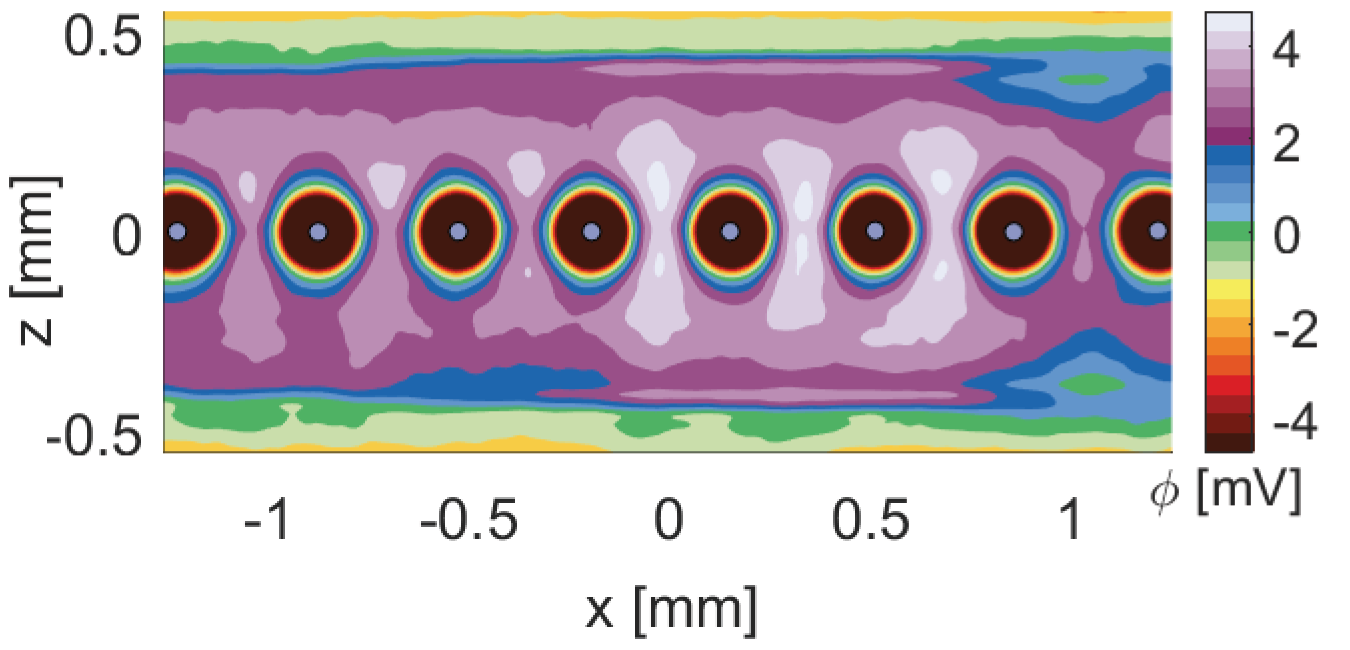}
    \caption{Electric potential in the vicinity of dust particles calculated by the DRIAD simulation. The simulation conditions closely reflect the 70 Pa 0.7 mA case.}
    \label{fig:DRIAD}
\end{figure}

We use the newly-founded q-Gaussian fits to calculate temperature in each domain using Eq. \ref{eq:qvariance}. Figure \ref{fig:temp gradient domains} a) $T_\|$ and b) $T_\perp$ show temperatures in units of Kelvin for 20 domains calculated for the 30 Pa 0.7 mA case. Each domain contained about 500 dust tracks, a minimum of 10 time frames per trajectory, which yields statistically significant amount of data for the velocity fits. The data indicates that temperatures gradients exist in the PK4 system. As seen from Eq. \ref{eq:duqv}, the temperature gradient and and the gradient in potential are related through the $q_v$ parameter. MD simulations of ions and dust in the PK-4, with conditions closely resembling the 70 Pa  0.7 mA case, reveal gradients in the electric potential surrounding each dust particles, as shown in Figure \ref{fig:DRIAD} \cite{Vermillion2022}. Examination of Figures \ref{fig:temp gradient domains} and \ref{fig:DRIAD} suggests that the dust clouds in PK-4 exhibit both temperature and electric potential gradients, in addition to q-Gaussian velocity distribution functions. This provides a qualitative argument for applying Eq. \ref{eq:qv}, which is dependent on the ratio of the thermophoretic force over the electric field force. Providing a quantitative validation for Eq. \ref{eq:duqv} is beyond the scope of the present paper and will be explored in future work.

\section{Discussion}
\label{sec:Discussion}

In this section, we compare of anomalous diffusion determined from fits to the MSD plots versus fits to the displacement histograms. We further discuss the possible physical mechanisms leading to the anisotropy observed when comparing the displacement and velocity distributions in the directions parallel and perpendicular to the electric field. Finally, we consider the nonequilibrium temperatures found from the velocity distribution fits.

\subsection{MSD and Displacement Histograms}

When analyzing the MSD plots fitting (Figure \ref{fig:msdalpha}), it seems like both $\alpha_\parallel$  and $\alpha_\perp$ increase with pressure for dc currents 0.35 mA and 0.7 mA. The trends are less obvious at 1 mA. These fits also indicate that none of the extracted $\alpha_\perp$ exponents correspond to a L\'{e}vy process, while all data for $\alpha_\parallel$ suggest a L\'{e}vy processes. Meanwhile, the $q_p$ coefficients extracted from the position histogram fits (Figure \ref{fig:qp}) also suggest that $q_{p\|}$ increases with pressure (i.e., enhanced superdiffusion), in agreement with the $\alpha_\parallel$ trends. However, only the highest pressure-currents case (70 Pa 1mA) in Figure \ref{fig:qp} indicates a L\'{e}vy processes (i.e., $q>5/3$). The $q_{p\perp}$ values extracted from single q-Gaussian fits show similar trends for the same currents as $\alpha_\perp$ but the 70Pa 1mA case suggests a L\'{e}vy process, unlike $\alpha_\perp$. The Bi-q-Gaussian fits yield $q_{p\perp2}$ values that exhibit closer similarity to the $\alpha_{\perp}$ trends. We can also see from Figure \ref{fig:q vs alpha error} that the fits for $q_{p\perp2}$ and $q_{p\perp1}$ yield lower percent error between the predicted and measured $\alpha_\perp$, strengthening the argument that the Bi-q-Gaussian is a more appropriate fit. The calculated diffusion constants (Figure \ref{fig:diffusion}) show an overall increase in parallel diffusion with pressure, while the perpendicular diffusion seems to decrease with pressure when a single q-Gaussian fit is used. These trends are not as clear for $D_{\perp1}$ and $D_{\perp2}$ which were obtained from a Bi-q-Gaussian fit. Both the  MSD fits and the displacement histogram fits indicate a pronounced difference in the diffusion regime when comparing the direction along the electric field versus the cross-field direction, which can be attributed to an anisotropy in the ion wake-mediated dust-dust interaction potential. 

\vspace{3mm}

The statistical analysis suggests that at high pressure-current conditions, a transition from superdiffusion to a L\'{e}vy process is expected in the axial direction. In the cross-field direction, the Bi-q-Gaussian fits suggest that the observed behavior is a superposition of a classical diffusion and superdiffusion, but no crossover to L\'{e}vy process is expected. These observations can be explained when considering the physical mechanisms causing the dust particles in the PK-4 experiments to organize into field-aligned filamentary structures. The alignment and the strong coupling of dust grains along the direction of the electric field is caused by enhancement of the ion wakefield focusing and elongation of the wakefield structure surrounding the dust grains (as discussed in \cite{Vermillion2022}). The formation of positive space charge due to ion focusing will cause a negatively charged dust grain to drift locally towards a nearby concentration of ions. This local drift is a nondeterministic microscopic effect, which is why it affects the microscopic motion by causing pronounced superdiffusion. The elongated ion wakefield structure surrounding the dust grains also causes a restoring force in the cross-field direction that keeps the individual dust particles aligned within the filament. This can explain the subdiffusive (or trapping) trends observed at small time scales in the cross-field MSD plots. Finally, if a dust particle escapes the confining potential that otherwise keeps it within a crystal-like filamentary structure, it cannot easily find force balance in between filaments. Instead, the dust particle makes large-scale jumps across the cloud until force balance is found within another filament. These jumps are frequently observed in video data from PK-4. 

\subsection{Histogram Shape in the Cross-field direction}
\label{subsec:PerpHist}

Both the position and velocity histograms obtained for the $\perp$ direction (Figure \ref{fig:qmsd}, \ref{fig:vhist}) were best fit by a bimodal distribution, i.e., the complex shape of the distribution can be viewed as a superposition of two simpler distributions. The single q-Gaussian which yielded the best results and fewest errors in the parallel direction, does not fit well in the perpendicular direction. We concluded that a Bi-q-Gaussian provided the best fit. Liu and Goree \cite{liu_particle_2018} showed that the dust velocity histogram in the PK-4 experiments can be described by what they called a Maxwellian 'core' and a Kappa distribution 'halo',

\begin{equation}
f_{M+\kappa}(v)=Ae^{\frac{-v^2}{v_th^2 }}+\frac{B}{(1+\frac{v^2}{\kappa v_{th}^2})^\kappa}.
\label{eq:core-halo}
\end{equation}

\begin{figure}[H]
    \centering
    \includegraphics[width=130mm,height=80mm]{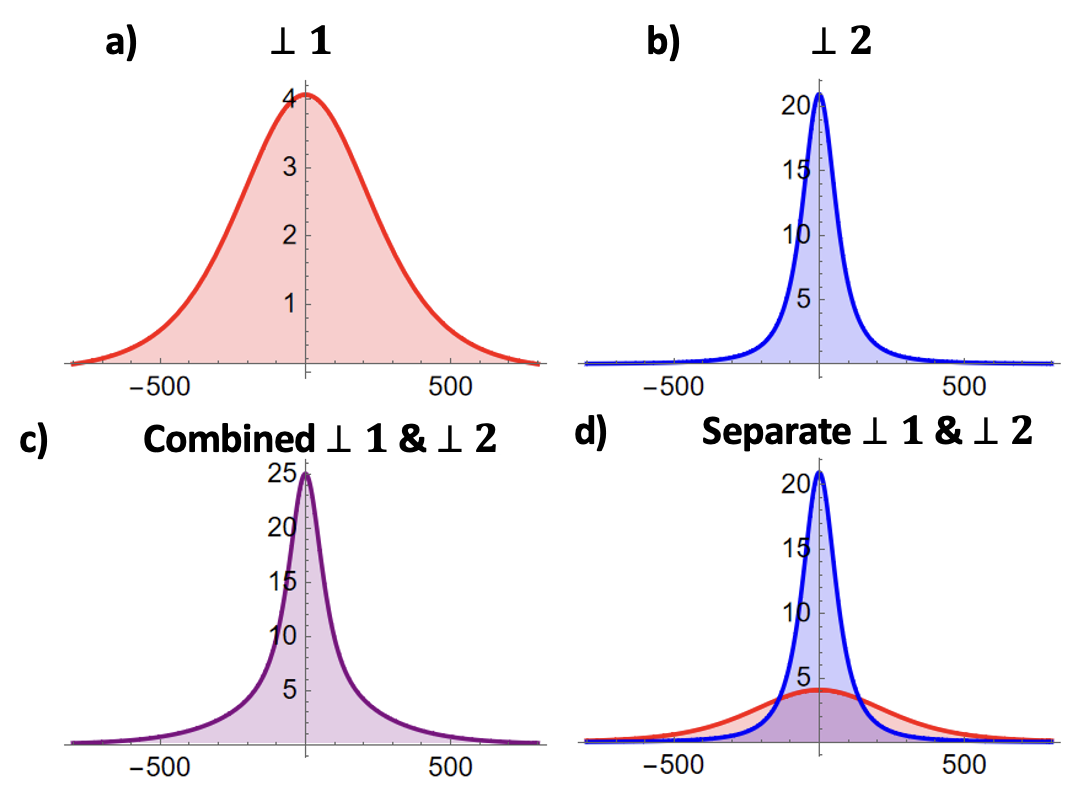}
    \caption{Components of the Bi-q-Gaussian illustrating two distinct populations: a) red $q_1=1.3$ and variance $\sigma_1 = 330$, while b) blue $q_2=1.63$ and $\sigma_2 = 63$. c) is the combined sum of the distributions in a) and b) while d) shows the same distributions overlapped but not added. These are the same as the $q_v$ values found for the cross-field velocity distribution in the 70 Pa 0.7 mA case. }
    \label{fig:decomp}
\end{figure}

Equation \ref{eq:core-halo} is very similar to the Bi-q-Gaussian as the Maxwellian 'core' is recovered by setting $q=1$ for one of the q-Gaussian distributions, while the Kappa 'halo' is identical to the q-distribution when one uses the known scaling relation $q=1+1/\kappa$ in the other q-Gaussian. In addition, Liu and Goree concluded that the cloud exhibited liquid-like coupling as evidenced by the calculated pair correlation function. The dust cloud in these experiments was confined in rf plasma produced by an rf coil and the observed cloud structure was homogeneous. In contrast, the present experiments were conducted in a pure dc discharge and the observed cloud structure was found to exhibit anisotropic coupling, suggesting both liquid and crystalline properties \cite{Baylor2024}. In both cases, the distributions consist of two terms, which implies a superposition of two distinct diffusion processes. 
Figure  \ref{fig:decomp} shows a decomposition of the Bi-q-Gaussian into individual q-Gaussians.

The superposition of the two q-Gaussian distributions shown in Figure \ref{fig:decomp} d) shows that one diffusion process (in red) is characterized by a small peak but large variance, while the other one (in blue) has both a large peak and large tails. The variance (and the corresponding temperature) of the blue distribution, however, is much smaller. Thus, the observed diffusion in the cross-field direction can be viewed as a superposition of two processes - one with higher temperature close to equilibrium and a second one with a lower temperature but farther from equilibrium.

\begin{figure}[H]
    \centering
    \includegraphics[width=140mm,height=50mm]{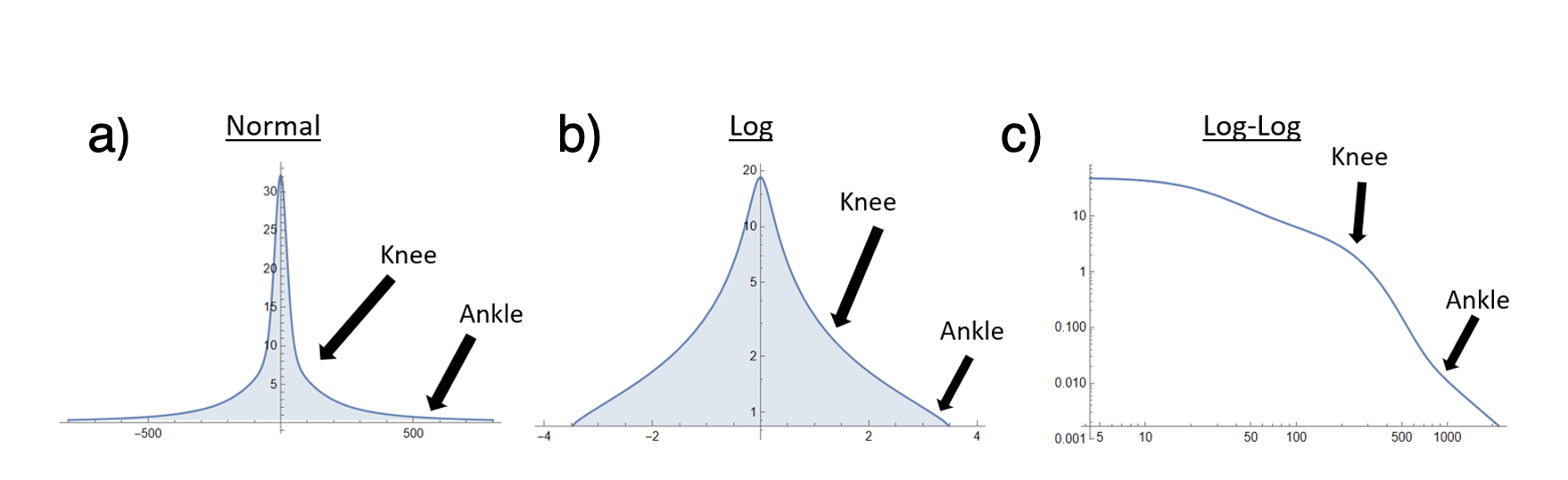}
    \caption{Analytical plot of the "knee and ankle" distribution a) with normal axis,b) a log scale, and c) log-log scale.}
    \label{fig:knee&ankle}
\end{figure}

This distribution shape is also called a 'knee and ankle' distribution, shown in Figure \ref{fig:knee&ankle}.

The 'knee and ankle' distribution has also been observed in solar wind ion and electron velocities and magnetic field fluctuations $\delta B$ \cite{treumann_stationary_2004,livadiotis_generation_2018,leubner_nonextensive_2005} as well as in fluxes of cosmic rays \cite{tsallis_fluxes_2003}. In \cite{leubner_nonextensive_2005}, a bi-kappa distribution was proposed to explain the scale-dependent changes in the solar wind PDFs obtained from spacecraft measurements. It was pointed out that small scale PDFs are highly non-Maxwellian, while Maxwellian is recovered for large scales. In PK-4, the dust particles seem to exhibit similar disparity between small and large scales as evident from the distribution fits. As the dust diffusion is highly dependent on ion-wake-mediated interaction potential, small time fluctuations in the ion wakefield can be the reason for the highly non-Maxwellian features in the PDFs. Such fluctuations are expected to result from high-frequency ionization waves that were recently discovered in PK-4 \cite{Hartmann2020}. 

\vspace{3mm}

In Space Physics, it is common to describe the non-Maxwellian velocity distributions using the kappa-distribution which is related to the q-Gaussian by $\kappa=\frac{1}{q-1}$. In \cite{livadiotis_generation_2018}, three physical mechanisms for generating a Kappa distribution were proposed: (i) Debye shielding, (ii) magnetic field biding, and (iii) temperature fluctuations. It was proposed that $\kappa$ has a negative correlation with the Debye number $N_D$ (number of particles in the Debye sphere). As $\kappa$ is inversely related to $q$, we expect a positive correlation between $N_D$ and $q$. In other words, increasing number of ions in the Debye spheres surrounding the dust should lead to enhancement of the superdiffusion. It was also proposed in \cite{livadiotis_generation_2018} that $\kappa$ is positively correlated with magnetic field fluctuations $\delta B$, which would imply negative correlation between $q$ and $\delta B$. While there is no magnetic field in PK-4, the charged dust particles are aligned to the external electric field. Thus, we conjecture that increase in electric field fluctuations should randomize the dust motion, thus, leading to classical diffusion. Finally, the negative correlation between $\kappa$ and the magnitude temperature fluctuations $\delta T^2$ proposed in \cite{livadiotis_generation_2018} suggests a positive correlation between $q$ and $\delta T^2$. The role of temperature and temperature fluctuations in PK-4 will be further discussed int he next section. According to \cite{Burlaga2005}, the magnetic field fluctuations, $\delta B$, nearer the Sun around $6.9$–$9.7$ AU, or even $43.6$–$47.2$ AU, remain in a non-Gaussian, meta-equilibrium state. However, farther out between $80$–$87$ AU, where the system has had more time ($\approx 1$ year) to relax, the magnetic field fluctuations reach a Gaussian distribution suggesting an equilibrium in magnetic field fluctuations. In other words, at these distances the solar-wind relaxes back to an equilibrium after going through a transition in the diffusion regime from $q > 5/3$ to $q < 5/3$ between $\sim 47$ AU and $\sim 80$ AU. Similar nonequilibrium-to-equilibrium regime transitions can be explored more easily and cost-effectively in laboratory settings with a dusty plasma experiment.

\vspace{3mm}

 \subsection{VDFs and Temperature}
 \label{subsec:vdfs}
 
As shown in Figure \ref{fig:Tq}, temperature decreases with increasing pressure for $T_\|$ and $T_\perp1$, aligning with the conclusions presented in \cite{Baylor2024}, where it was seen that dust 'cooled off' due to the increased role in dust neutral collision, which was calculated by use of pair-correlation functions. The temperatures $T_\perp2$, however, seem to show the opposite trend of increasing with increased pressure. The Tsallis coefficients $q_{v\parallel}$, never exhibit values greater than $5/3$, which suggests that there is no L\'{e}vy processes for this direction. We also notice that $q_{v\parallel}$ tends to decrease with increasing pressure (thus, improving equilibrium), while increasing current slightly increases $q_{v\parallel}$ (thus, driving the system away from equilibrium). This is reasonable as an increase in pressure causes an increase in neutral collisions, meaning that the random collision process brings the dust motion to equilibrium. Meanwhile, the current provides an electric field which affects the ion streaming and focusing surrounding the dust, which can cause local attractive interactions among dust particles and macroscopic regions of positive space charge. 

\vspace{3mm}

While the coefficients $q_{v\perp}$ and $q_{v\perp2}$ suggest L\'{e}vy processes, the trends are not as simple to interpret. To visualize this, we look at Figure \ref{fig:decomp} a), which shows a hot  population, more so in equilibrium, while the cold, shown in blue, has much lower kinetic energy. This is most likely due to the trapping potential that creates the observed filamentary structures within the dust cloud. However, the blue population is far from equilibrium. This population is confined in the cross-field direction and much of its kinetic energy is in the parallel direction. When this energy transitions to the cross-field direction, it may do so suddenly, resulting in the larger tails of the distribution, or big jumps of the dust particles in space. In other words, Figure \ref{fig:decomp} illustrates a one population that is highly peaked with significant tails (kurtosis), giving a small variance, i.e., temperature, and another distinct population that exhibits high variance but less kurtosis. Based on this interpretation, increasing pressure appears to drive the first population $q_{\perp1}$ away from equilibrium, while improving equilibrium for the second population $q_{\perp2}$ for all but the 1 mA case. The confining forces keep the dust particles in a crystal-like structure, mostly aligned in the parallel direction, but these forces do not always succeed in maintaining this configuration. Similar to slipping out between tight tweezers, particles that escape these confining forces become energetic in the perpendicular direction. Once the particles escape, their movement is primarily dictated by the ion wakefield caused by the electric field, leading to high parallel diffusion. In other words, the forces confining the dust must act in the parallel direction; otherwise, the particles would accelerate due to the electric field. 

\vspace{3mm}

An examination of the domain velocity distributions at 0.7 mA reveals that $q_{v\perp}$ tends to be higher on the edges of the cloud and smaller in the center at low pressure, but larger on the top and smaller on the bottom for 46 Pa, 70 Pa. We also see that for 30 Pa at 0.7 mA there exist clear gradients in both the $T_\parallel$ and $T_\perp$. (We remind the reader that there is only one $T_\perp$ here since a single q-Gaussian fits well the individual domain data in the cross-field direction.) Unexpected high dust temperature has been a point of contention in the dusty plasma community because the experiments are always in a room temperature setting, however this is better understood when considering all forms of energy in strongly coupled systems. We propose that, in addition to the kinetic temperature of the dust, the additional apparent energy comes from variation of the electrostatic potential. Typically one calculates temperature as $E=k_BT=1/2mv_{th}^2$. However, we propose $E+U=k_BT+Q_{dust}\langle\Delta\phi_{float}^2\rangle$, meaning that room temperature of the dust particles can be recovered by employing the following expression

\begin{equation}
    T=\frac{1}{2}\frac{m}{k_B}v_{th}^2 - \frac{e}{k_B}\langle N_{dust} \rangle\langle\Delta\phi_{float}^2\rangle.
\end{equation}

An order of magnitude calculation using this equation reveals that variation in the floating electric potential on the order of $~10^{-5}$ Volts could explain temperature discrepancy. This conclusion is explained more rigorously in \cite{Avinash2011}.

\subsection{Energy Dissipation}

The energy dissipation and thermal state of PK-4 dusty plasma clouds was studied in detail by McCabe, et al. \cite{mccabe_energy_2025}. It was found that the onset of polarity switching of the electric field in PK-4 (used for trapping the cloud in the camera's field of view) causes an initial expansion in the dust cloud leading to an increase in thermal energy. Simulation results from that paper suggest that the onset of polarity switching causes a modification in the effective dust screening length due to reconfiguration of the dust during the expansion. The present work uses the same set of PK-4 experiments discussed in McCabe, et al., which allows for a meaningful comparison of results across the two studies. An important distinction is that McCabe, et al. used data obtained at and immediately after the onset of polarity switching, while the analysis here was performed on data collected later in the experiment. However, one of the main findings in McCabe, et al. is that an extended time (much greater than the Epstein drag decay) is required to dissipate thermal energy in these dusty plasma clouds. Thus, it is reasonable to expect that the clouds do not have enough time to come to an equilibrium state for the duration of these experiments. 

\vspace{3mm}

This is confirmed by the statistical analysis presented here. Specifically, the velocity distributions for all cases were found to be non-Maxwellian with high energy tails, which is evident from the $q_v > 1$ shown in Fig. \ref{fig:q_nonequilibrium}. Here we further build on the study by McCabe et al. by identifying differences in the non-Maxwellian distributions obtained for velocities parallel versus perpendicular to the direction of the electric field. Parallel to the direction of the field, the $q_{v\parallel}$ values are only slightly greater than $1$, which suggests that the thermalization, or energy dissipation, along this direction should be close to classical. However, in the perpendicular direction, the $q_{v\perp}$ values suggest a crossover from a superdiffusive to a L\'{e}vy process. In other words, a small sub-population of the particles has an order of magnitude higher energy than the bulk particles, which leads to inefficient energy dissipation. Physically, the polarity switching of the electric field causes a modification of the ion wakefield surrounding the dust grains, which in turn leads to the observed self-organization of the dust particles into field-aligned chains. If a dust particle escapes a chain, it is likely to experience big jumps in the cross-field direction until it finds force balance as part of another chain. These jumps are clearly observable in the video data.

\vspace{3mm}

In addition to the onset of polarity switching, changes in the discharge conditions are another mechanism that can cause anisotropic interaction potential and slow energy dissipation in the PK-4 dusty plasma clouds. Hartmann, et al. \cite{hartmann_inhomogeneities_2020} used a 2D particle-in-cell with Monte Carlo collisions discharge simulation to investigate inhomogeneities in the PK-4 plasma. It was found that, on $\mu s$-scale, the structure of the positive column is characterized by the propagation of ionization waves. This prediction was experimentally validated in the PK-4 replica at Baylor University, though for pressure and current conditions slightly higher than those considered in this study. In the presence of ionization waves, the electric field and ion/electron densities can reach amplitudes up to 10 times larger than their average values with important implications to the ion-wakefield mediated dust-dust interaction potential. Numerical work by Matthews, et al. \cite{matthews_ionization_2020} and Vermillion, et al. \cite{vermillion_temporal_2022} showed that the propagation of ionization waves through the dust can modify the streaming velocity of ions and introduce time-dependent anisotropies in the ion wakefield structure surrounding each grain. 

\vspace{3mm}

The statistical analysis conducted in this work provides quantitative experimental evidence for the presence of these anisotropies in the PK-4 ISS dusty plasmas. For all pressure-current conditions, the histogram shapes, nonextensive parameters $q_p$ and $q_v$, diffusion coefficients, and temperatures show differences when calculated for the direction along the electric field versus the cross-field direction. In addition, this anisotropy on the diffusion seems to become more pronounced at higher pressures. This can be inferred from single distribution fits of the diffusion in Figure  \ref{fig:diffusion} a) and b) in the $0.35 mA$ and the $1 mA$ case, where $D_{\parallel}$ increases with pressure increase, while $D_{\perp}$ decreases with pressure increase. Physically, the general decrease in most diffusion values in the transverse direction at higher pressure coincides with the observation that the dust alignment in filaments longitudinally improves at higher pressure, which has been confirmed by calculating pair correlation functions \cite{Kost_Proceedings_2023}. As the filamentary structures form, the interparticle separation and coupling strength change with direction, which in turn changes the flow of thermal energy and energy dissipation in these clouds.  However, these trends are not exactly followed by the $0.7 mA$ case, most notably, the $46.1 Pa$ data point. While there may be particle tracking errors in the absolute particle positions data, we believe that the statistical treatment using position displacements and fits to histograms minimizes this source of errors. In addition, a measure of goodness of fit using $R^2$ and $\chi^2$ normalized by degrees of freedom, for all cases, including the $0.7 mA-46.1 Pa$, is typically around $R^2\approx 0.99$ for both parallel (single q-Gaussian Figure \ref{fig:diffusion} a) and perpendicular (single q-Gaussian for Figure \ref{fig:diffusion} b and bi-q-Gaussian for Fig. \ref{fig:diffusion} c and d) distribution fits. The parallel direction distribution fits have normalized $\chi^2<10^1$ while normalized $\chi^2\approx10^2$ for single q-Gaussian perpendicular fits and $\chi^2\approx10^1$ in the bi-q-Gaussian fits, which is expected due to better fitting of the tail behavior. Thus instead of fitting error, the probable source of the disagreement in the $46.1 Pa$ $0.7 mA$ data point is the difference in dust densities for different plasma conditions.  Re-examination of the data from the PK-4 plasma glow camera, we could see that for all currents, the $46.1 Pa$ experiments had one large cloud at the center of the discharge tube, while for the $28.5Pa$ and the $70.5 Pa$ cases, there were multiple smaller dust clouds forming around plasma striations throughout the discharge tube. Since the number of particle dispenser shakes was kept constant, it is reasonable to expect that the difference in size and number of clouds will lead to a difference in dust density, which in turn, affects the calculated statistical properties. In a follow-up experiment (not discussed here), the plasma conditions were kept constant, while the dust density of the same cloud was varied through cloud compressions. These results will be a subject of a follow-up paper.

\vspace{3mm}

It has been demonstrated numerically \cite{matthews_ionization_2020} that the formation of filamentary structures in PK-4 is enhanced by the presence of ionization waves. Since the PK-4 experiment on the ISS is not equipped with a high-speed video camera, we do not have direct experimental evidence that these ionization waves occur for all examined pressure-current conditions. However, the statistical analysis presented here provides a way to quantify the possible effect of ionization waves through the observed anisotropies in the dust diffusion. The present findings will benefit from comparison to further discharge simulations and experimental studies using the on-ground PK-4 replicas  to better understand the parameter space where the ionization waves should be expected to dominate. 

\section{Conclusions}
\label{sec:Conclusions}

The presented analysis of of PK-4 dusty plasma experiments highlights the features of anomalous dust diffusion resulting from an interplay between temperature fluctuations and anisotropies in the dust-dust interaction potential caused by the application of external electric field to this strongly coupled complex system. Notably, we used non-extensive Tsallis statistics to quantify the nonequilibrium state of the system in the directions along and across the electric field using q-Gaussian distributions. The nonequilibrium parameters $q_p$ and $q_v$ (extracted from position and velocity distributions) have distinctly different values for the directions parallel and perpendicular to the electric field, as shown in Table \ref{tb:fitparams}. Particle motion parallel to the electric field is superdiffusive and for several pressure-current cases, superdiffusion crosses over to a L\'{e}vy process. In the cross-field direction, the diffusion is again anomalous, but the bi-Gaussian fit best describing the histograms suggests a superposition of two processes. One of these processes is similar to classical diffusion, while the other one is superdiffusive, but not L\'{e}vy. We attribute the directional dependence of diffusion to anisotropies in the dust-dust interaction potential and the dust temperature throughout the cloud. The former is caused by the anisotropic shape of the ion wakefield clouds surrounding the dust grains, which leads to the alignment of the dust particles into filaments. Inside a filament, macroscopic regions of positive space charge form along the direction of the electric field due to ion focusing causing local dust accelerations towards these attractors. Similarly, the cross-field motion of the dust is restrained due to a restoring force from the streaming ions within the wakefield structure. However, if a dust particle escapes the filament, it will experience large jumps across the cloud until it finds force balance within another filament (which is visible in the video data). We hypothesize that the higher parallel diffusion causes more energy exchange between particles, thus bringing the system to equilibrium. This is seen by $q_{v\parallel}$ being only slightly greater than $1$ while the parallel diffusion is large. Also, $q_{v\perp}$ is relatively greater than $1$ and the perpendicular diffusion is much smaller. This is also supported by the trends that exist between the parallel diffusion and $q_{v\parallel}$ with increasing pressure, which are inversely proportional. The same is true of perpendicular diffusion and $q_{v\perp}$ also being inversely proportional, which may suggest that increased diffusion contributes to increasing the equilibrium.

\vspace{3mm}

In addition, the kinetic temperature is found to vary throughout the dust cloud as shown in Figure \ref{fig:temp gradient domains}. Analysis of subdomains within the dust clouds reveals that a single q-Gaussian fit works well even in the cross-field direction for small domain sizes. Since the $q_v$ extracted from velocity histograms quantifies the nonequilibrium state of the system, the domain analysis allows for a nuanced understanding of cloud properties, including the structural anisotropy, diffusion, and equilibrium state. Specifically, we conclude that the different domains within the same dust cloud are characterized by different diffusion regimes driving the global state away from equilibrium. Temperature gradients were identified for the dust in the PK-4. While all the temperatures were higher than room temperature, we propose that the observed high kinetic temperature is due to electrostatic fluctuations. Finally, we conclude that dusty plasma systems, such as those in PK-4, are excellent for studying nonequilibrium systems, anomalous diffusion and the physical origins of the two phenomena. As the mathematical description of these processes is universal, we expect that the dusty plasma experiments can be a practical means to investigate other nonequilibrium complex systems such as the solar wind.

\section{Acknowledgments}

This material is based on work supported by NSF grant numbers 2308742,  2308743, EPSCoR FTPP OIA2148653, 1903450, and 1740203, NASA grant number 80NSSC21K0381. All authors gratefully acknowledge the joint ESA - Roscosmos ``Experiment Plasmakristall-4'' onboard the International Space Station. The microgravity research is funded by the space administration of the Deutsches Zentrum für Luft- und Raumfahrt eV with funds from the federal ministry for economy and technology according to a resolution of the Deutscher Bundestag under Grants No. 50WM1441 and No. 50WM2044 

\section{Data Availability}
The data supporting the findings of this study were obtained from the PK-4 experiment conducted on the International Space Station, a collaborative project between the European Space Agency (ESA) and Roscosmos. Due to international agreements and data sharing policies between ESA and Roscosmos, the data are not publicly available. However, they can be made available from the corresponding author upon reasonable request and with permission from the collaborating agencies.

\bibliographystyle{unsrt}

\end{document}